\documentclass[fleqn,usenatbib]{mnras}

\usepackage[T1]{fontenc}

\DeclareRobustCommand{\VAN}[3]{#2}
\let\VANthebibliography\thebibliography
\def\thebibliography{\DeclareRobustCommand{\VAN}[3]{##3}\VANthebibliography}

\usepackage{newtxtext,newtxmath}

\usepackage{graphicx}	
\usepackage{amsmath}	
\usepackage{booktabs}

\title[Chemodynamics of Milky Way disks]{Disentanglement of the chemodynamical assembly: mapping the Milky Way disks}

\author[E. Cantelli et. al.]{
Elvis Cantelli,$^{1}$\thanks{E-mail: elvis.cantelli@usp.br}
Ramachrisna Teixeira,$^{1}$
\\
$^{1}$Instituto de Astronomia, Geofísica e Ciências atmosféricas - Universidade de São Paulo, Rua do matão, 1226, 05508-090, São Paulo, Brazil \\}

\date{Accepted 2024 April 12. Received 2024 April 06; in original form 2023 September 30}

\pubyear{2024}

\begin{document}
\label{firstpage}
\pagerange{\pageref{firstpage}--\pageref{lastpage}}
\maketitle

\begin{abstract}

The formation and structure of the Milky Way has a fundamental role in our understanding of the universe and its evolution, and thanks to the Gaia mission and large spectroscopic surveys, we live an exceptional moment of data availability, allowing us to trace the building blocks of the Galactic disk and their relations.
In this sense, we propose here the exploration of a large dataset in a top-down fashion, elaborating a similarity map of the local Galactic volume in order to segregate and characterise its main components, searching for hints about their relations.
We have used the t-SNE algorithm with chemical, orbital and kinematic properties of the stars to produce 2D manifolds and dissect their structure by isolating populations to further analyse their behaviour.
The young thin disk could be clearly separated from the older thick disk, also showing a puzzling transition zone with hints about the aftermath of the Gaia-Sausage-Enceladus merger.
Moving groups and resonant features also appear prominently in the maps, splitting the disk into inner and outer portions as consequence of the resonances produced by the Galactic bar.
The dynamical halo appears as an extreme end related to the heated portion of the thick disk, showing sub-structures corresponding to known accreted populations.
Open and globular clusters also appear in their chemical/evolutionary context.
We present details of the developed strategy, an overview of the different populations and their relations, as well as a discussion and insights of our results in the scenario of Galactic evolution.

\end{abstract}




\section{Introduction}

A richer and more detailed understanding of the history of our Galaxy has, since the 18th century, been one of the main goals of astronomy, with early research conducted by William Herschel's maps of star densities, spiral structures detected by William Parsons in other galaxies, the flattened disk shape proposed by Jacobus Kapteyn, followed by the great debate between Shapley and Curtis that finally culminated in \cite{hubble1925} confirming the "island universes" scenario in the 1920's. 
Our view about its nature and structure started to take shape with \cite{lindblad1926} and \cite{oort1927} deducing the rotation of the Galaxy, \cite{baade1944} separating the stellar populations I and II, \cite{baade1951} inferring the existence of the spiral arms, and \cite{vyssotsky1951} and \cite{schwarzschild1952} discovering the Galactic halo. 
More recently, a bar \citep{deVaucouleurs1964} and an older thicker disk \citep{gilmorereid1983} came to our knowledge, as well as several stream-like structures (eg. \citealt{ibata1994}, \citealt{helmi1999}) that trace the hierarchical building of the $\Lambda$-CDM \citep{Springel_2006} cosmological scenario, and evidences (eg. \citealt{nissen2010}, \citealt{deason2013}) of a major merger event that built this configuration, as proposed earlier in \cite{chiappini1997}.
As we expand and refine the databases that supports our understanding of the Milky Way, new details emerge that lead us on to continue scrutinising our Galaxy.

With the advent of the Gaia mission \citep{gaia_main} and large spectroscopic surveys, the data availability became significantly larger, encompassing a more isotropic coverage and thus, more homogeneous sampling of the local volume of the Galaxy, allowing the discovery of many new components of the Galactic halo, such as the Gaia-Sausage-Enceladus (hereafter GSE, \citealt{belokurov2018}, \citealt{haywood2018}, \citealt{helmi2018}), Sequoia \citep{Myeong_2019} and several others (see \citealt{Naidu_2020}, \citealt{Forbes_2020}, \citealt{Myeong_2022}).

These halo substructures, especially the GSE, are associated with earlier dwarf galaxy merger events, which have shown to be the direct cause of the heating of an early Galactic disk and formation of the "hot" thick disk, previously assigned as an "inner halo" \citep{DiMatteo_2019} or a "metal-weak thick disk" (eg. \citealt{Beers_2002} and \citealt{Hayes_2018}).
Still, a proper characterisation of disk populations is the target of a myriad of studies (eg. \citealt{Gilmore_1989}, \citealt{rama_2007}, \citealt{Hawkins_2015}), employing distinct techniques to segregate the populations based on their kinematics, chemistry and ages and disentangle these components, tracing their contribution as Galactic building blocks, as well as dynamically-induced populations by bar resonances (see \cite{drimmel_2023}) such as the Hercules stream \citep{PerezVillegas_2017}.
The disk heating in plane-perpendicular velocity beyond $\sim$10 kpc was associated by \cite{das2024} with a first encounter with the Sagittarius dSph satellite galaxy.
\cite{Feuillet_2022} were able to identify old, accreted populations within the dynamically cool stellar disk, unveiling a whole new range of disk components yet to be discovered.

The use of unsupervised learning algorithms is rapidly growing among astronomers as an efficient method to classify and characterise observed and simulated data in a vast and diverse range of applications. 
The dimensionality-reduction algorithm t-distributed stochastic neighbour embedding (t-SNE, \citealt{tsne}) is one of the most useful algorithms to visualise and characterise high-dimensional data, already employed in several areas, including star-forming galaxies \citep{Steinhardt_2020}, stellar spectroscopy \citep{Traven_2017}, galaxy morphology \citep{dai2018visualizing}, and even instrumentation \citep{House_2023}.

\cite{anders_2018} employed the t-SNE for a sample of 530 HARPS-GTO stars with reliable chemical abundances to separate different Galactic components in a multi-dimensional chemistry space, and was able to separate the disc sub-populations in the solar vicinity with more reliable visualisations than common 2D chemical diagrams.
\cite{da_Silva_2023} has recently used t-SNE to analyse 6618 metal-poor stars with a space composed of orbital actions and chemical abundances, enabling the characterisation of Galactic halo groups and their connections to the major accretion events (Sequoia and GSE).
\cite{Queiroz_2023} applied the t-SNE to samples of sub-giants of three large spectroscopic surveys (4638 for APOGEE, 9420 for GALAH, and 12834 for LAMOST) by using a chemical and age parameter space, separating the thin disk from the thick disk by clustering the manifold via HDBSCAN.
In this work we propose to use both chemical abundances and orbital parameters to feed t-SNE with stars in the local volume focused on the disk, without relying in random initialisation, and for a much larger sample of stars.

Our aim is to characterise the different Galactic components with sub-samples of two high-resolution spectroscopic surveys, namely APOGEE (Apache-POint Galactic Evolution Experiment, \citealt{apogee}) and GALAH (GALactic Archaeology with HERMES, \citealt{desilva2015}) using kinematic parameters calculated from the Gaia catalogue (orbital energies, angular momenta, radii, eccentricities and velocities, since these properties can summarise the orbital profiles) and different sets of chemical abundance ratios (metallicity, alpha-element, odd-Z, iron-peak and s-process, summarising the most important chemical-tagging features). 
We assume that this chemodynamic parameter space is capable of describing the formation and evolution of the main Milky Way populations and enable their segregation within a large area around the sun.
For this, we apply the t-SNE algorithm to perform the dimensionality reduction to a 2-D space and produce a structural map, where each chemo-kinematic population accommodates in separated locuses from each other, decomposing and bringing into light the several sub-structures that lie hidden in the Galactic disk to dissect each one and assemble a comprehensive scenario of the local disk configuration.
In this sense, it is expected to obtain a clear separation between thin and thick disk based on their different chemical signature and velocity patterns, signatures of resonant features from the Galactic bar and their associated kinematic populations, clues on older populations lying in the outermost portion of the disk, as well as the isolation of dynamical halo stars and their possible connection to the thick disk.
In our strategy, instead of trying to isolate manifold regions only by their concentration or using clustering algorithms directly, we colour the maps with the parameter values by using different palettes that can visually bring out the map behaviour and guide the segregation process more conceptually.

The paper is outlined as follows.
In Section \ref{data_methods} we describe the dataset and sample selection, and in Section \ref{orbital} we give details on the orbital integration and show validations of the Galactic potential used.
The t-SNE strategy and parameters for manifold convergence are shown in Section \ref{tsnesec}, the results with the component segregation are shown in Section \ref{results}, and the analysis of the different Galactic components and their properties is shown in Section \ref{discussion}.
Finally, in Section \ref{conclusion} we summarise the results and bring new insights about the use of the technique proposed in this work

\section{Data}
\label{data_methods}

\subsection{Survey data}

We have obtained the data from Gaia DR3 \citep{gaia_dr3} catalogue by applying the selection criteria described in \cite{gaiaHR_2018}, requiring a determined value for radial velocity and a magnitude in G band up to 15, where DR3 radial velocity sample is essentially complete. 
This primary Gaia DR3 sample had 20 million stars with full 6D phase space. 

The distances were calculated using the formalism of \cite{abj2018} through the code made available by F. Anders at https://github.com/fjaellet/abj2016, with 0.3 pc for distance PDF resolution.

The spectroscopic survey data were obtained from APOGEE DR17 (\citealt{apogeedr17}, \citealt{apogeespec}, \citealt{sloan}) and GALAH DR3 \citep{galah_dr3}
by downloading the full catalogues available in the websites of the surveys.

APOGEE is a high-resolution ($R\approx22500$) spectroscopic survey in the near-infrared operated primarily in the northern hemisphere with the Sloan telescope \citep{Gunn_2006} and later in the southern hemisphere with the duPont telescope \citep{Bowen_73}, with a dedicated spectrograph \citep{Wilson_2019}.
It was designed to observe red clump stars, red giant branch stars, RR-Lyrae, among many other stellar objects.
The fields are sub-divided into two types: grid pointings which sample a semi-regular grid in Galactic longitude and latitude over a component of the Milky Way (such as the disk and bulge), and non-grid pointings that sample particular objects of interest (dwarf galaxies, tidal streams, clusters).
The stellar parameters and elemental abundances are obtained via the ASPCAP automated pipeline \citep{GarciaPerez_2016}, which compares the observed spectra to libraries of synthetic spectra calculated with model atmospheres (described in detail in \citealt{Jonsson_2020}) and spectral line lists \citep{Smith_2021}.

GALAH is a high-resolution ($R\approx28000$) spectroscopic survey in the optical region operated in the southern hemisphere with the Anglo-Australian Telescope's HERMES spectrograph \citep{Sheinis_2015}.
The target selection had two phases: the first one being a magnitude-limited survey covering stars with magnitude between $12 < V < 15$, declination $\delta<10^{\circ}$ and angular Galactic plane distance $|b|>10^{\circ}$; and the second one more focused on age estimation, sampling the main-sequence turnoff, and the Galactic plane previously not covered.
In its third data release, the stellar parameters and abundances were obtained with Spectroscopy Made Easy \citep{Piskunov_2016}, where the main stellar parameters are estimated by fitting synthetic spectra to 46 segments containing lines with reliable data for this analysis, improving the result further in a second iteration.
The elemental abundances are obtained separately for each element, using the same segment-driven strategy for the parameters.
A full description is given in \cite{galah_dr3}.

The cross-match of the surveys with the primary Gaia sample by their \textit{source\_id} resulted in a reduction of 733901 to 342263 stars for APOGEE and 588571 to 491790 for GALAH.
The sharp reduction was due to the selection function of Gaia, that required a measured radial velocity and $G_{mag}$ < 15 (giving a little extra from the completeness drop of the radial velocity around G = 14.5, \citealt{Katz_2023}).
After this, we applied a selection filter to the [Fe/H] and [Mg/Fe] abundances, requiring the errors to be < 0.2 dex with no pipeline flags, resulting in a reduction to 307871 stars for APOGEE and 391710 for GALAH (see Table \ref{errors}).
Although the radial velocities from the spectroscopic surveys have smaller errors due to the higher resolution, we chose to use the Gaia RVS to keep the homogeneity of the sample phase space.

\begin{table}
\caption{Median, 5$^{th}$ and 95$^{th}$ percentiles of the error distributions of parallaxes, proper motions (pm), radial velocities (RV), effectivce temperatures (T$_{eff}$), gravities, abundances and signal-to-noise ratios in the main-sequence (MS) and red giant branch (RGB) subsamples for APOGEE and GALAH.}
\begin{center}
\begin{tabular}{lcccccc}
\hline\hline
 & \multicolumn{3}{c}{APOGEE} & \multicolumn{3}{c}{GALAH} \\   
\cmidrule(lr){2-4} \cmidrule(lr){5-7}
Parameter & p05 & p50 & p95 & p05 & p50 & p95 \\
\hline
 & \multicolumn{6}{c}{Main sequence} \\
parallax [mas] & 0.01 & 0.018 & 0.039 & 0.01 & 0.015 & 0.03 \\
pm$_{\alpha}$ [mas/yr] & 0.01 & 0.018 & 0.042 & 0.01 & 0.016 & 0.032 \\
pm$_{\delta}$ [mas/yr] & 0.011 & 0.016 & 0.037 & 0.01 & 0.014 & 0.029 \\
RV [km/s] & 0.3 & 1.53 & 5.74 & 0.5 & 1.94 & 5.15 \\
T$_{eff}$ [K] & 9 & 20 & 54 & 75 & 94 & 123 \\
log G [dex] & 0.017 & 0.025 & 0.045 & 0.178 & 0.183 & 0.195 \\\relax
[Fe/H] & 0.005 & 0.009 & 0.016 & 0.052 & 0.084 & 0.132 \\\relax
[Mg/Fe] & 0.011 & 0.016 & 0.025 & 0.037 & 0.094 & 0.181 \\
SNR & 42 & 126 & 397 & 22 & 42 & 101 \\
\hline
& \multicolumn{6}{c}{Red giant branch} \\
parallax [mas] & 0.01 & 0.015 & 0.025 & 0.01 & 0.015 & 0.024 \\
pm$_{\alpha}$ [mas/yr] & 0.009 & 0.015 & 0.028 & 0.009 & 0.015 & 0.026 \\
pm$_{\delta}$ [mas/yr] & 0.01 & 0.014 & 0.023 & 0.01 & 0.013 & 0.022 \\
RV [km/s] & 0.16 & 0.61 & 2.93 & 0.18 & 0.72 & 2.81 \\
T$_{eff}$ [K] & 6 & 9 & 18 & 74 & 93 & 130 \\
log G [dex] & 0.02 & 0.025 & 0.038 & 0.185 & 0.206 & 0.281 \\\relax
[Fe/H] & 0.007 & 0.008 & 0.013 & 0.046 & 0.071 & 0.114 \\\relax
[Mg/Fe] & 0.009 & 0.012 & 0.021 & 0.034 & 0.075 & 0.172 \\
SNR & 77 & 219 & 703 & 26 & 55 & 135 \\
\hline\hline
\end{tabular}
\end{center}
\label{errors}
\end{table}

\begin{figure}
	\includegraphics[width=1\columnwidth]{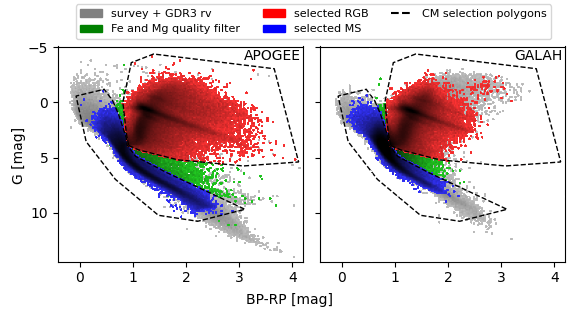}
        \includegraphics[width=1\columnwidth]{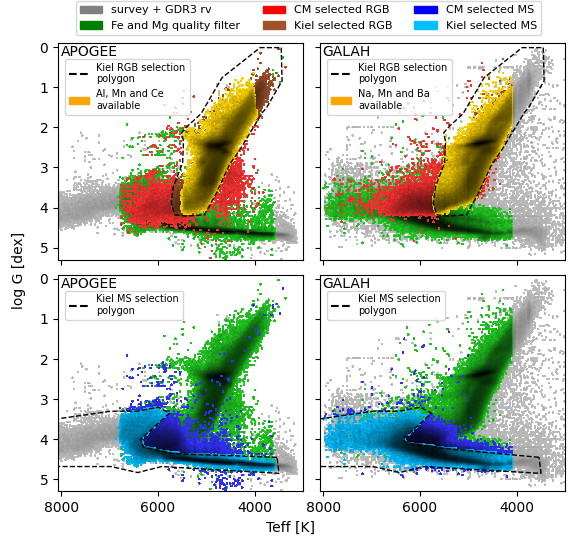}
    \caption{Selection cuts of the main-sequence and giants sub-samples from the APOGEE (left) and GALAH (right) samples. Upper panels show the Gaia cross-correlation in gray, filtered stars by Fe and Mg quality in green, and the selected RGB (red) and MS stars (blue) as subsets of quality-filtered stars. Middle panels show the Kiel diagram with the RGB selected by CM (red), the selection by the Kiel diagram cut (brown), and the stars that have determined abundances for the elements discussed in Section \ref{elemental} in yellow. Same for the lower panels, but for the MS selected by CM (blue) and Kiel diagram cut (light blue). All polygons used for selection cuts are displayed by dashed lines.}
    \label{CM-cut}
\end{figure}

We separated the red giant branch (RGB) based on the criteria from \cite{katz2018} for stars with absolute magnitude in G band $M_{G} < 3.9$ and intrinsic colour $(G_{BP}-G_{RP})_{0} > 0.95$, but we chose to perform a visual cut in the colour-magnitude (CM) diagram to compensate for the reddening of sub-giants above $G_{M} > 4$, and also because there were spurious main-sequence (MS) stars in the upper portion of the Kiel diagrams of the samples, preventing a direct log G selection cut.
Furthermore, reddening data were not available for the full sample and, even if applied for the available stars, produced a glitchy CM diagram.
Figure \ref{CM-cut} displays the selection cuts for MS and RGB stars from the APOGEE (left) and GALAH (right) datasets. 
The dashed polygons represent the cuts, with colour-coded points to distinguish between the successive subsets produced by filters and cuts. 
The coordinates for the polygons are shown in the Appendix \ref{poly_append}.
In the upper panels we show the CM diagram of Gaia cross-correlation (gray) with the Fe and Mg filter (green), and the subsequent selection cuts for RGB (red sub-samples) and MS (blue sub-samples).
This CM selection was projected into the Kiel diagram to perform further cuts avoiding the turn-off and contamination from the MS as shown in the middle panels for the RGB (brown sub-samples), and lower panels for the MS (light blue sub-samples), resulting in 155773 and 132468 stars on the RGB sample, and 99037 and 161668 stars on the MS sample for APOGEE and GALAH respectively.
All selections were made respecting the same boundary polygon on both surveys, except for GALAH MS kiel cut with two vertexes modified to admit more stars.
The last step was to impose that the RGB samples had determined abundances for the elements used for t-SNE inputs, as discussed in Section \ref{elemental}, where we achieved a final RGB catalogue (yellow sub-samples) of 151534 stars for APOGEE and 124686 for GALAH, shown in the middle panels.

The main analysis and t-SNE maps were made with the RGB sample, since these stars have more reliable radial velocities, proper motions and chemical abundances, where the typical errors are smaller than the MS with larger signal-to-noise ratios, as shown in Table \ref{errors}.
We display the median errors for the selected samples of each survey together with the 16th and 84th percentiles, since the error distributions are mostly asymmetric.
Also, APOGEE has roughly $\sim$300 stars in the MS with all the required elements in the filtered sample, enabling only the use of the RGB for both surveys to keep sampling consistency.
The GALAH main-sequence sample was used in the Galactic potential validation for the orbital parameters described in Section \ref{orbital}, since this sample encompasses around $\sim$1.2 kpc around the sun, in contrast with about $\sim$0.6 kpc from the APOGEE MS sample.
Moreover, GALAH MS sample is much larger and includes hotter and more massive younger MS stars, as seen in the lower panels of Figure \ref{CM-cut}, which captures more of the thin disk kinematic behaviour.

\begin{figure}
	\includegraphics[width=0.9\columnwidth]{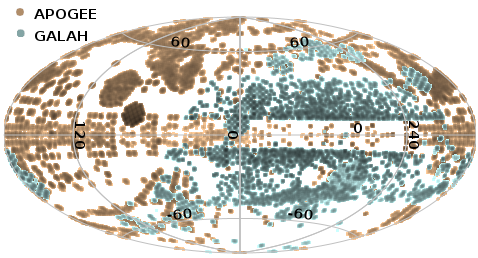}
    \caption{Coverage of the APOGEE (ocre) and GALAH (cyan) RGB final samples in galactic coordinates.}
    \label{aitoff_map}
\end{figure}

The distribution of the final sample RGB stars is shown in the projection of Figure \ref{aitoff_map}, with the sampling space mostly dependent on APOGEE DR17 and GALAH DR3.
In this visualisation we can notice the southern celestial hemisphere sampling of GALAH in contrast with the northern-dominated sampling of APOGEE, and also a lack of stars in the Galactic plane spanning roughly $\sim$180 degrees towards the Galactic centre.

\subsection{Elemental abundances}
\label{elemental}

If we are segregating different populations in the Galaxy by tracing astrophysical events such as a major merger or accretion of dwarf galaxies, the inclusion of more chemical data (in addition to metallicity and $\alpha$-elements) is very useful since mass inflows can cause starburst events in-situ, and the own satellite abundances give hints about their mass and evolutionary history.
In particular the odd-Z elements, the iron-peak and the heavy elements, since associations of light elements like Na and Al together with Mn and Mg are able to segregate the halo/accreted populations (\citealt{Hawkins_2015}, \citealt{Das2020}, \citealt{Feuillet_2021}), while the s- and r-process elements can trace different astrophysical scenarios, older populations and halo stars (\citealt{Aguado_2021}, \citealt{Carrillo_2022}, and \citealt{da_Silva_2023}), as well as serving as good chemical clocks (\citealt{Spina_2017}, \citealt{Magrini_2018}, see Appendix \ref{ages}).

\cite{Nandakumar_2022} has produced calibration catalogues to scale between APOGEE and GALAH abundances, but since our elemental selection spans beyond [Fe/H] and [$\alpha$/Fe] with different species representing each elemental group (Al as light and Ce as heavy for APOGEE, and Na as light and Ba as heavy for GALAH), we chose to use the original abundances.
Moreover, the use of different chemical catalogues in the same segregation methodology can be a good validation to the results produced by each one.

\begin{figure}
	\includegraphics[width=\columnwidth]{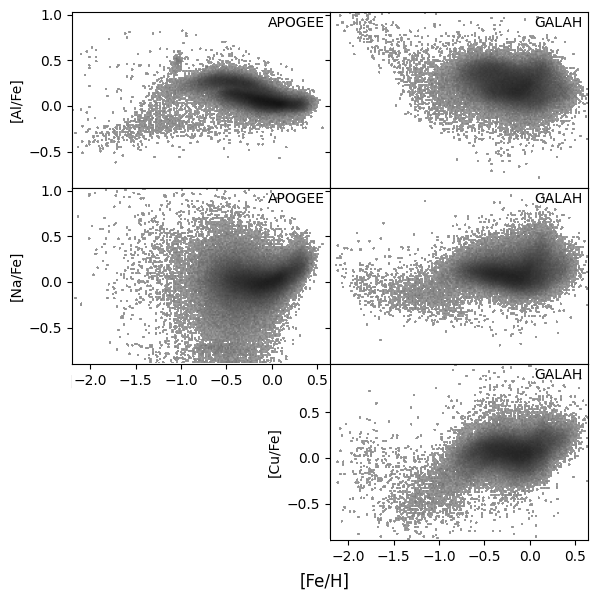}
	\includegraphics[width=\columnwidth]{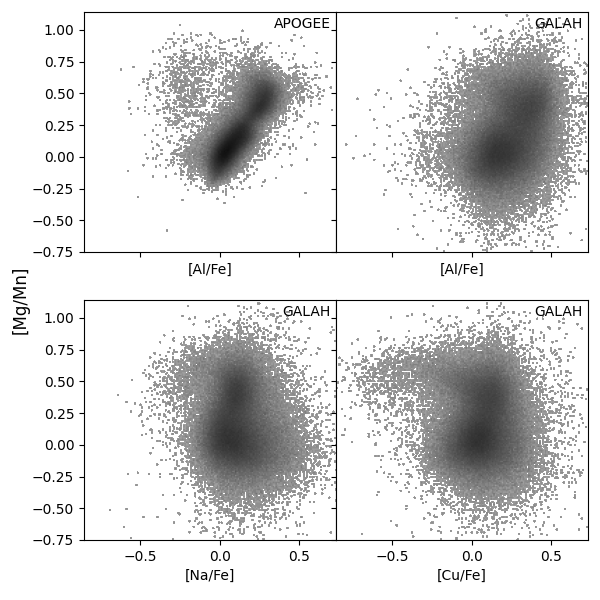}
    \caption{Elemental abundance patterns in the RGB sub-samples of spectroscopic surveys. Top panels: odd-Z elements vs [Fe/H], where the sparse "tail" represents the dynamical halo with accreted populations, for APOGEE on the left and GALAH on the right. Lower panels: [Mg/Mn] vs odd-Z, showing the separability of the accreted populations in different cases for the surveys.}
    \label{Na_Al_comparison}
\end{figure}

The main challenge in this abundance scenario is to properly select the available elements within each survey, as each one has a different set of elements beyond [Fe/H] and [Mg/Fe]. 
We show in Figure \ref{Na_Al_comparison} the odd-Z and Mn elements from both surveys in order to compare their quality and usability to detect accreted populations.

In APOGEE, the Al is well determined and shows the low [Al/Fe], [Mg/Mn] abundances that characterise low star-formation rates typical of low-mass galaxies as the indicative of accreted stars, where in GALAH this population is over-estimated for Al and can be seen better with Na and Cu (see eg. \citealt{Giribaldi_2023}). 
\cite{Kobayashi_2020} shows that these odd-Z elements have similar trends with metallicity, and for the GALAH sample, Cu seems to produce a better distinction of the accreted population than Na (lower right panel of Figure \ref{Na_Al_comparison}).
However, even though Cu has a similar nucleosynthetic origin, it is more related to the iron-peak than to light elements like Al and Na, and this use for Cu is shown in the Figure 27 of \cite{galah_dr3} in the plane [Mg/Cu] vs [Na/Fe] as an alternative to [Mg/Mn] vs [Al/Fe] to segregate the accreted populations.
Moreover, APOGEE lists Cu for the abundance list, but the whole column is nulled, preventing its use for the analysis.
Even though this plane is able to segregate the accreted halo, the disk collapses to a central concentration at [Na/Fe] $\sim$ 0.1 and [Mg/Cu] $\sim$ 0, whereas in the [Mg/Mn] vs [odd-Z/Fe] planes of Figure \ref{Na_Al_comparison} there is a split of two main concentrations, one around solar value for both ratios that is associated with the metal-rich thin disk, and the other at [Al/Fe] $\sim$ 0.25 and [Mg/Mn] $\sim$ 0.5 that is associated with the thick disk (discussed furthermore and shown in Figure \ref{apogee_dyn_halo}).
Therefore, we chose the use of Mn as the iron-peak element and Na as the light odd-Z for the GALAH analysis and preserve some consistency with APOGGE.

The iron-peak elements are mainly produced during the thermonuclear explosion of type-Ia supernovae, and in this case, Mn is produced more than Fe \citep{Kobayashi_2009}, where \cite{Hawkins_2015} showed that the Mn abundance can be used to distinguish between thin and thick disk stars by tracing the type-Ia supernovae enriched population.

\begin{figure}
	\includegraphics[width=\columnwidth]{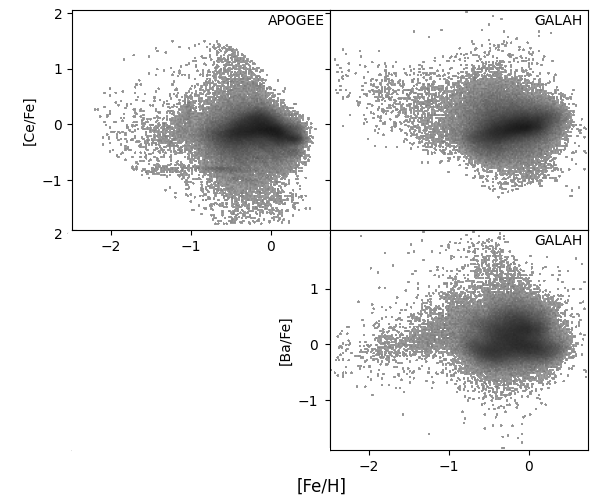}
    \caption{Abundances of heavy s-process elements vs [Fe/H] for the RGB sub-samples of the surveys. The monotonic pattern of GALAH [Ce/Fe] does not match with APOGEE [Ce/Fe], where GALAH [Ba/Fe] ratio shows a much better agreement.}
    \label{Ce_Ba_comparison}
\end{figure}

For the heavy elements, the only one available in APOGEE is Ce, which is an s-process (second peak) element. In the GALAH sample, the [Ce/Fe] - [Fe/H] distribution is very poor in comparison to the expected trends of \cite{Kobayashi_2020} and the APOGEE counterpart, as shown in Figure \ref{Ce_Ba_comparison}. 
The closest correspondent element in the GALAH sample is Ba, which has a more similar distribution and has the same nucleosynthetic origin.

\subsection{Ages}

In the process of segregating and characterising populations, the stellar ages can play a major role in helping to determine the sequence of events that took place in an astrophysical scenario, and with these parameters, it is possible to separate populations that would be indistinguishable for sharing similar properties in other variables and establish a timeline of the assembly of the Galaxy. 

The GALAH sample has ages estimated from the bayesian BSTEP \citep{Sharma_2017} code, based on stellar observed parameters and the PARSEC-COLIBRI isochrones \citep{Marigo_2017}. 
In the other hand, APOGEE ages were estimated using the astroNN \citep{Leung_2018} trained with the asteroseismic ages from \cite{Montalb_n_2021} and \cite{Mackereth_2019}.
However, there are some inconsistencies between the two age estimations if we compare them by using the common stars between the two surveys, as discussed in Appendix \ref{ages}.
In face of these large inhomogeneities between the default age determinations of APOGEE and GALAH (astroNN and BSTEP), and the lack of a large enough matching sample from other sources such as StarHorse \citep{Queiroz_2023}, we chose not to use ages as a main input parameter for t-SNE, as it would introduce more errors and unwanted effects in an already large dimensional parameter space, leaving the age as a validation parameter.
Instead, we chose to use the [Ce/H] and [Ba/H] ratios that, as discussed in Appendix \ref{ages}, may also work as an age estimator better than [Fe/H].

\subsection{Orbital parameters}
\label{orbital}

As already stated by \cite{antoja-helmi}, the Gaia era showed that assuming a time-independent, axisymmetric potential in dynamical equilibrium for the Milky Way is definitively incorrect, requiring a more rich approach to extract orbital parameters in order to better characterise stellar kinematics.
To calculate the orbit trajectories and orbital parameters, we employed the \textit{galpy} code \citep{Bovy_2015} using the modified \textit{MWPotential2014} with halo mass $M_{h} = 1.4 \times 10^{12} M_{\odot}$, twice the original value, together with a bar using the \textit{DehnenBarPotential} parametrised with $\Omega_{b}=$ 39 km/s/kpc, $R_{\Omega}=$ 3.5 kpc, and azimuth angle of 28$\deg$ from the Sun-Galatic centre line, with the sun at $R=8.3$ kpc, following the results of \cite{Lucey_2023} and solar motion from \cite{schonrich2012}. 
The integration was made backwards in time with 1000 steps down to -800 Myr from present, which is roughly equivalent to 5 bar periods.
The parameters obtained were the apocentric and pericentric radii $R_{ap}$ and $R_{per}$, eccentricity $e$, maximum Galactic plane distance $Z_{max}$, orbital energy $E$, total and Z-axis angular momenta $L$ and $L_{z}$, galactocentric radial, tangential and vertical velocities $V_{r}$, $V_{\phi}$, $V_{z}$, as well as mean orbital radius $\Bar{R}_{o}$.

The use of a consistent bar potential is necessary in order to reproduce and better constrain kinematic populations that originate from the bar resonances, such as Hercules, Hyades and Sirius (\citealt{Khoperskov_2022}, \citealt{PerezVillegas_2017}) which are well-established moving groups.
The constraining of a connection between the resonances and their associated moving groups could validate the possible kinematic effects that may arise from this barred structure and its influence on the outer portions and dispersion-dominated components such as the thick disk.

\begin{figure*}
\centering
  \includegraphics[width=0.45\linewidth]{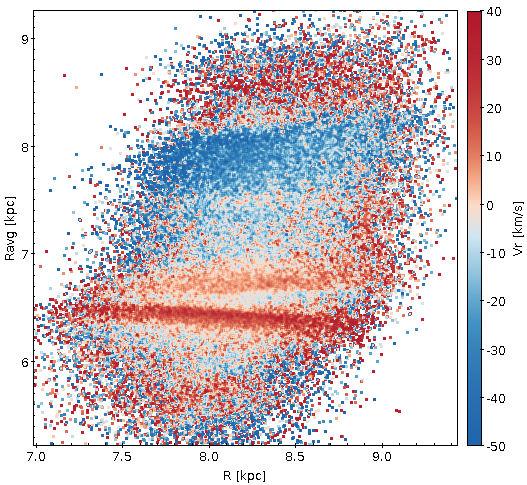}
  \includegraphics[width=0.41\linewidth]{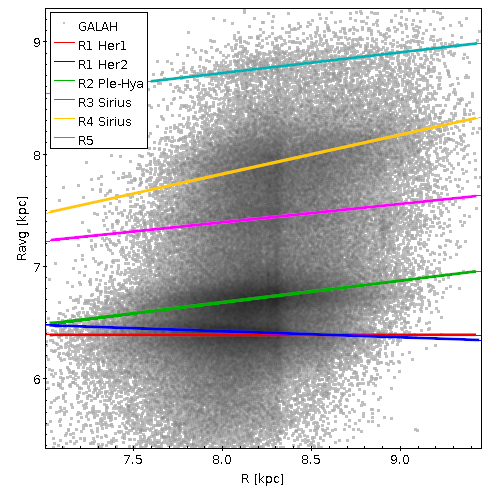} \hfill
  \includegraphics[width=0.46\linewidth]{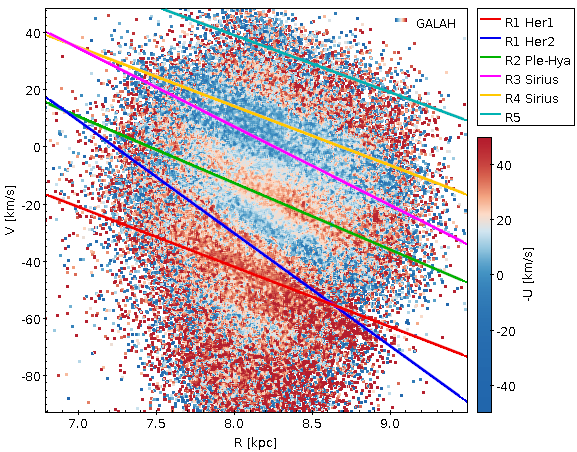}
  \includegraphics[width=0.42\linewidth]{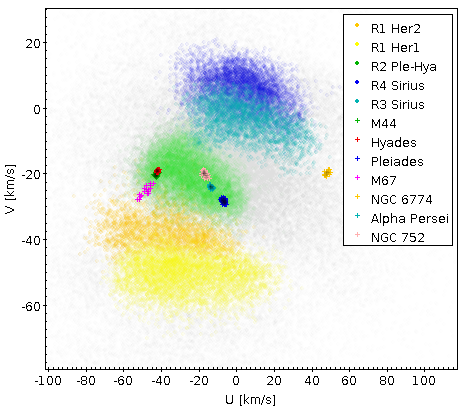}
  \caption{Radius and velocity planes for the GALAH sub-sample. Galactocentric radial velocity ridges are shown in the upper left panel, where the corresponding star density ridges (R1 to R5) induced by $\Bar{R}_{o}$ accumulation are shown with guide lines in the upper right panel. The ridges selected from the upper panels are projected in the bottom left panel as the guide lines, and the corresponding stars of the ridge fiducial selection are shown in the U-V plane together with open clusters in the bottom right panel.}
\label{orbit_ridges}
\end{figure*}

In order to verify the Galactic radius of the bar resonances that arise from the calculated orbits and match their expected positions, we used the GALAH MS sub-sample (with a much larger size than the APOGEE counterpart) to inspect the ridges produced by the accumulation of $\Bar{R}_{o}$ (Figure \ref{orbit_ridges}), since the MS sample contains the younger and bluer stars, thus it is more suitable to investigate the rotation-dominated thin disk with the dynamical effects produced by the presence of a bar.

We made fiducial selection cuts around the density ridges (R1 to R5) that appear along the $\Bar{R}_{o}$ which are also visible as $V_{r}$ ridges in the upper panels of Figure \ref{orbit_ridges}, and by fitting linear trends to these ridges it is possible to marginally constrain the location of these features in other planes, such as the V-R plane colour-coded by -U \citep{Chen_2022}.
In this visualisation, we projected the fiducial lines corresponding to each selected density ridge group, enabling the comparison of the orbital effects of the bar potential (mainly $\Bar{R}_{o}$) with directly observable parameters (R, U, V), where the density ridges of $\Bar{R}_{o}$ match the location of -U velocity ridges. 
This qualitative agreement is an indication that the bar parameters employed in the potential are a good representation of the true Galactic bar potential, since higher pattern speeds shifts the Lindblad resonances (and consequently the position of $\Bar{R}_{o}$ ridges) towards the Galactic centre, dissociating the -U velocity ridges from the $\Bar{R}_{o}$ density ridges.
Moreover, these fiducial selections are shown in Figure \ref{orbit_ridges} in the U-V plane together with stars from open clusters available in our sample in positions, matching the locations of the moving groups of Hercules, Hyades and Sirius, with the ridge positions in a good agreement with those detected by \cite{Bernet2022}.
The red and royal blue lines, once a single density ridge (R1) at $\Bar{R}_{o}\sim6.5$ kpc (R1 Her1), were splitted with the help of V-R plane, where the high $-U$ value around -40 km/s in $V$ associated with this ridge shows an additional small ridge (R1 Her2) with a more pronounced slope.
This duplicity may be first two components (Hercules 1 and 2) of the multimodal structure of the Hercules stream shown by \cite{Asano_2020}.

\section{t-SNE}
\label{tsnesec}

The t-distributed stochastic neighbour embedding (t-SNE) algorithm \citep{tsne} is a non-linear dimensionality reduction technique for visualising high-dimensional data in a low-dimensional space, modelling each object in a two- or three-dimensional point where similar groups are reduced to nearby clusters, thus creating islands and gaps in such a way that, for our purpose, results in a 2-D space where distinct Galactic components are expected to appear clustered in distinct regions.
A good tutorial to understand the mechanics on how t-SNE works is outlined in \cite{tds_tsne}, but in a nutshell, it calculates the conditional probabilities with gaussian distribution (that can be understood as similarities) for the data points in the large-dimensional space, performs a first guess of the low-dimensional space by using PCA (this is for our case, but it could be set to random) and calculates the conditional probabilities with a Student-t distribution.
After this, it iterates over the low dimensional space to minimise the Kullback-Leibler divergence as the cost-function via gradient descent and stops at a defined number of iterations or when the divergence reaches no significant change in the last iterations.
As a result, it returns the points arranged in the low dimensional space that can be seen as a similarity space, where similar groups clump together and relate to neighbouring groups.

\cite{anders_2018} showed a default parametrisation for the \textit{sklearn} implementation of t-SNE where two of the main three parameters (\textit{learning\_rate}, hereafter \textit{lr} and \textit{early\_exaggeration}, hereafter \textit{ee}) should be set to 1 and $0.1 \cdot N_{points}$, respectively, that worked for a sample of 530 stars.
The \textit{learning\_rate} governs how fast the gradient descent changes the positions of the samples in the 2D space across iterations, and the \textit{ee} governs how fast the clusters of data move across the space in the initial phase of the convergence, or as defined in the \textit{openTSNE} \citep{open_tsne} documentation, it increases the "attractive forces" between points and make clusters more compressed.
The \textit{perplexity} (hereafter \textit{pp}) is the parameter that adjust the bandwidth of the gaussian kernels centered in each sample which will admit the points to calculate similarities, and can be thought as an analogous to the \textit{k} nearest neighbours.
In practice, it governs the preservation of structures, where large values preserve global structures and smaller values preserve local structures, since it regulates the amount of neighbours "seen" by a point to be comprehended in what the distribution admits as a local cluster.

In preliminary tests, having our sample with significantly more stars, the \textit{ee} parameter showed to perform better around $0.005 \cdot N_{points}$, and \textit{lr} with values ranging between 600 and 800, as the maps should resemble scattered clouds instead of entangled tapes or uniform balls of points, following the examples of \textit{sklearn} documentation, where values in the range 100-150 were appropriate to keep both global and local structures.
Another good tutorial with several examples on how the convergence behave based on the perplexity are available in \cite{wattenberg2016how}.
\cite{Belkina_2019} showed that raising the number of iterations in the early exaggeration phase led to better accommodations of global structures, where clumps belonging to the same group tended not to scatter around the final manifold.
We also experimented with modifying the \textit{sklearn} implementation to allow the input of early exaggeration \textit{n\_iter} to perform more tests, achieving slightly better results, but as discussed further on, it contributed to our choice of using another implementation of t-SNE.

The \textit{sklearn} implementation has several limitations in comparison to the more recent implementation of \textit{openTSNE}, including speed, \textit{early\_exaggeration} iterations (natively, instead of modifying the source code), different \textit{exaggeration} settings for each phase, the possibility to obtain snapshots along the iterations via callback functions (enabling the inspection of the convergence), as well as novel features, such as perplexity annealing (the modification of perplexity along the run) and the additiion of new data to an existing embedding.
A first Galactic component classification can be used as a control along t-SNE iterations and different parametrisations (as discussed further in Section \ref{results-manifold}), using different variable spaces and subsample selections, enabling consistency checks of the manifolds produced. 
Having the power to fine-tune the parameters and customise the embedding, it is possible to primarily separate large structures (such as thin disk, thick disk and halo) with a large perplexity, then sub-classify them by reducing the perplexity to continue the iterations, resulting in an embedding that makes a better representation of the Galactic components based on the hierarchy of parameters.
A description of the parameters and procedures used to fine-tune the t-SNE and the convergence maps will be presented in Section \ref{results-manifold}.

\subsection{Manifold convergence}
\label{results-manifold}

The process of convergence for the t-SNE implied in exhaustive parametrisation tests in order to obtain a satisfactory iteration process in which we could observe the gradual separation of our pre-labeled structures by their relational hierarchy. 
For example, one could expect that the alpha-rich, dynamically hot component (namely the canonical thick disk) would separate early from the young metal-rich thin disk, and subsequently show some other structures such as the dynamical halo.
With the \textit{openTSNE} ability to change and tune hyperparameters in a continuous iteration process and to communicate with the iterator object via a callback function, we were able to build a parameter cascade for each one of the RGB datasets (APOGEE and GALAH) and obtain our final manifolds, detailed in the following sections.

The kinematic and chemical parameters to use as t-SNE features were $E$, $Z_{max}$, $L$, $L_{z}/L$, $V_{\phi}$, $e$, $R_{ap}$, $R_{per}$, [Fe/H], [Mg/Fe], a light odd-Z element ([Na/Fe] for GALAH and [Al/Fe] for APOGEE), [Mn/Fe] and a heavy s-element ([Ba/Fe] for GALAH and [Ce/Fe] for APOGEE).
The maximum Galactic plane distance $Z_{max}$ is extremely useful to dynamically trace the thin and thick disks, as well as the halo.
We introduce $L_{z}/L$ as a direct measure of orbital inclination, based on how much the Z component (associated to the Galactic plane orbital motion) contributes to the total angular momentum, where smaller values approaches perpendicular orbits and negative values easily represents retrograde motion.
$E$ and $L_{z}$ are typically used in the Lindblad diagram, and are able to separate accreted and dispersion-dominated populations, and even though we include only $L$, $L_{z}$ is recovered by the use of $L_{z}/L$, with $V_{\phi}$ not only reinforcing its character, but also being able to segregate the well-known moving groups in the U-V plane (as shown in Figure \ref{movingroups-manifold} furthermore).
Here we use the instant angular momenta and eccentricities, since these variables are not conservative in a non-axisymmetric potential.
The apocentric and pericentric radii place the constraints on how deep a star may dive to the centre or elongate to the outermost parts of the Galactic disk, and the eccentricity $e$ is fundamental to trace circular or elongated orbits.
One could argue that $R_{ap}$, $R_{per}$ would suffice to map this orbital feature, but as the euclidean space is linear, $e$ cannot be obtained as a linear combination of the radii, which is not the case for $L_{z}$, that it is linearly dependent of $L_{z}/L$ and $L$.
We performed a min-max normalisation on all parameters (except $L_{z}/L$ and $e$ that are already in the 0-1 range) before executing the t-SNE runs, a very important step not only for t-SNE, but also for machine learning algorithms in general.

\begin{figure*}
    \centering
    \includegraphics[width=\linewidth]{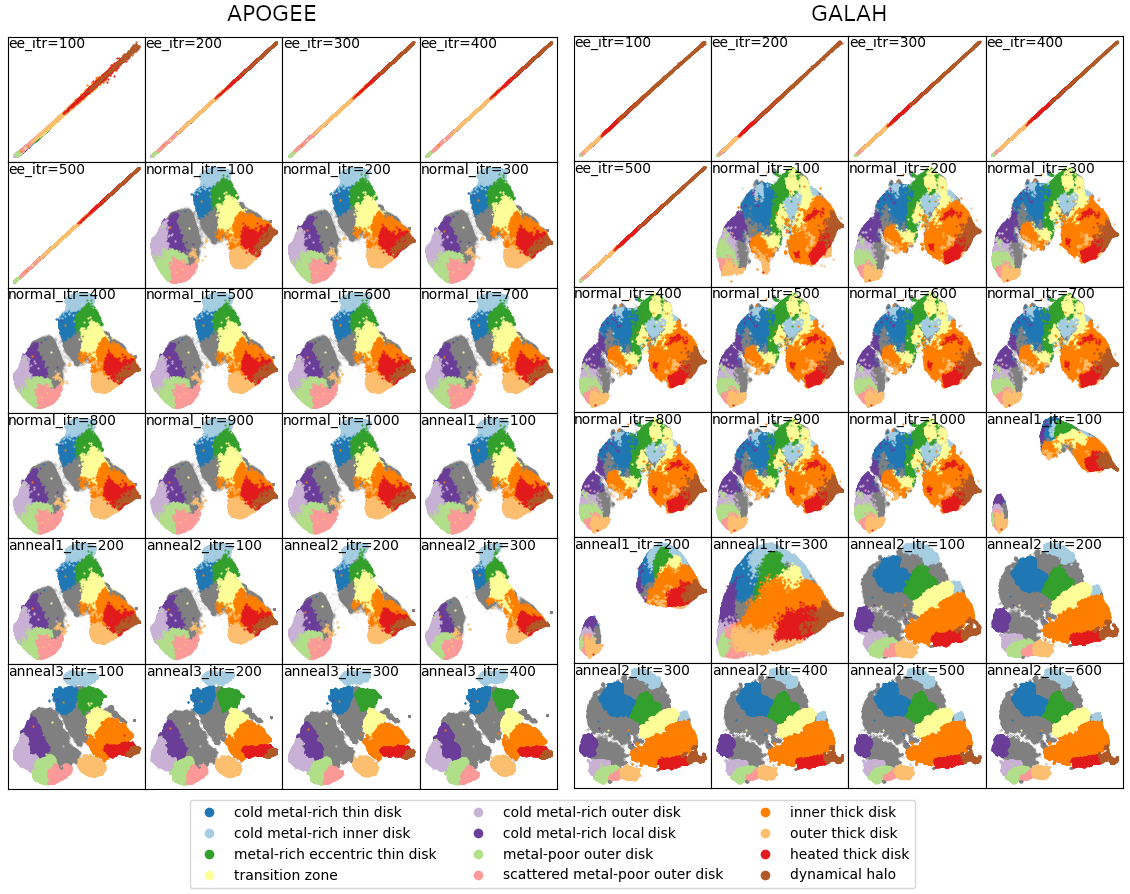}
    \caption{t-SNE convergence iterations with colour-coded reference mapping for the APOGEE sample in the left panel and for GALAH sample in the right panel. The named regions were chosen in an iterative process while converging the manifold. We show the final reference classification made by visually comparing, identifying and cross-matching both APOGEE and GALAH data. The last annealing iterations are shown only up to 600 since the manifold achieves a state that is very close to its final shape. The iteration stages are annotated in the subplots, representing one at every 100 iterations, according to the scheme of Table \ref{tSNE_hyperpars}, and they follow a left-to-right and top-to-down order.}
    \label{tsne_convergence}
\end{figure*}

The initialisation was set to "pca", where the convergence is not affected by a random state, producing a deterministic result that sustains itself for any run, depending only on the convergence parameters.
The metric used was the cosine distance, as recommended by \textit{openTSNE} documentation.
The cascade used for APOGEE consisted in a primary t-SNE parametrisation with \textit{early exaggeration} (corresponding to the first 5 and subsequent 10 tiles of Figure \ref{tsne_convergence}), followed by two annealing steps ($anneal_1$ and $anneal_2$ on Figure \ref{tsne_convergence}), and a final annealing ($anneal_3$), both without \textit{early exaggeration}.
For GALAH, we made two steps both with \textit{early exaggeration}, with the first one corresponding to the same pattern as APOGEE in Figure \ref{tsne_convergence}, and the second labeled as $anneal_1$ for the \textit{early exaggeration} phase and $anneal_2$ for the normal iterations phase.
The hyperparameters used in each convergence step is summarised in Table \ref{tSNE_hyperpars}.

\begin{table}
\centering
\caption{Hyperparameters \textit{early exaggeration} (ee), \textit{exaggeration} (ex), \textit{learning rate} (lr), \textit{perplexity} (pp), and iterations used in each convergence step for APOGEE and GALAH.}
\label{tSNE_hyperpars}
\begin{tabular}{lcccccc} 
\hline\hline
Step & \textit{ee} & \textit{ex} & \textit{lr} & \textit{pp} & \textit{ee} iter. & normal iter. \\
\hline
 & \multicolumn{6}{c}{APOGEE} \\
primary     & 80 & 2 & auto & 250 & 500 & 1000 \\
annealing 1 & - & 2 & 90 & 170 & - & 200 \\
annealing 2 & - & 5 & 70 & 130 & - & 300 \\
annealing 3 & - & 1.6 & auto & 50 & - & 2000 \\
\hline
 & \multicolumn{6}{c}{GALAH} \\
primary   & 90 & 1.7 & auto & 150 & 500 & 1000 \\
annealing & 9 & 1.8 & auto & 25 & 300 & 2000 \\
\hline\hline
\end{tabular}
\end{table}

Following this relational hierarchy idea, we traced the manifold convergence by colouring and labelling reference groups for some well-known Galactic components (thick disk, metal-rich thin disk and dynamical halo for example), monitoring the displacement behaviour of the structures in the manifold in an iterative process, as shown in Figure \ref{tsne_convergence} for the two survey samples.
The identification of the reference groups was done by visually inspecting parameter-coloured maps (as shown for example in Figures \ref{apogee_variables} and \ref{galah_variables}, discussed in detail on Section \ref{results}), and matching these groups between APOGEE and GALAH in such a way that both manifolds showed the same structures with very similar chemical and kinematic patterns.

\begin{figure*}
    \centering
    \includegraphics[width=\linewidth]{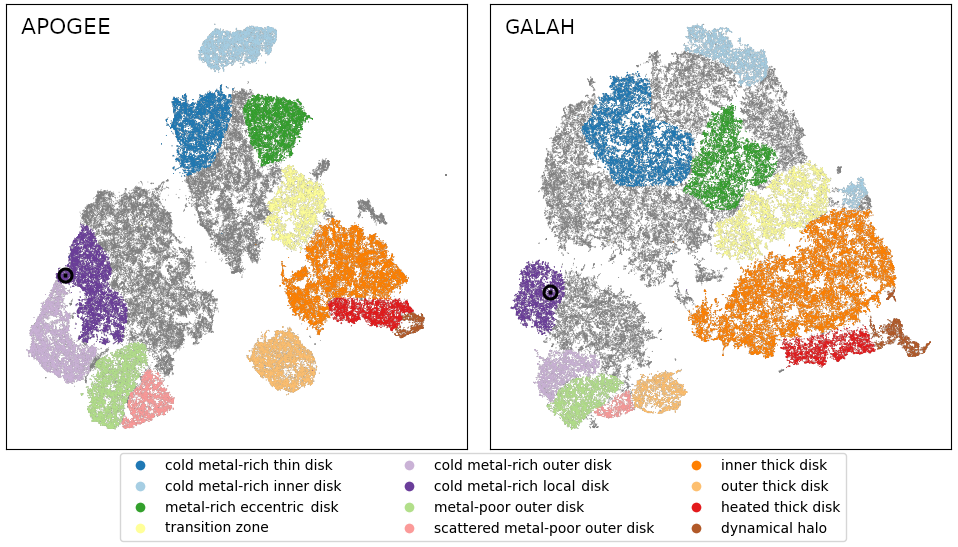}
    \caption{Final manifolds achieved For APOGEE (left panel) and GALAH (right panel). The $\odot$ mark represents the sun location in the manifold. The axis ticks were omitted since the manifold coordinates have no physical meaning. The labelled "inner" groups are for $R_{avg} < 7$ kpc and "outer" are for $R_{avg} > 7$ kpc.}
    \label{final_manifolds}
\end{figure*}

The splitting of the blobs change at the start of each annealing, especially for GALAH, and we see that the identified groups tend to accommodate and define themselves more when the new \textit{early exaggeration} starts in the annealing stage.
The light orange blob representing the \textit{outer thick disk} transits between the orange portion (inner part of the thick disk) and the lower left portion that represents the outer populations, and show some relation to the \textit{scattered metal-poor outer disk}, discussed in more detail at Section \ref{thindsec}.

\section{Results}
\label{results}

The main goal of this work is to produce a map where we can visualise the Galactic structure in an intuitive fashion, enabling us to spot sub-structures and relations that would not be easily accessible otherwise.
The definition of regions were done visually, and can be extremely useful to dissect some specific populations, their peculiarities, and how they fit in the Galactic assembly, as already shown in the previous section.

The final manifolds are shown in Figure \ref{final_manifolds} with some of the main identified structures labelled in different colours for both APOGEE and GALAH with their mapped positions coinciding between the surveys. 
The legend is grouped by the large clusters to guide the eye.

\begin{figure*}
    \centering
    \includegraphics[width=\linewidth]{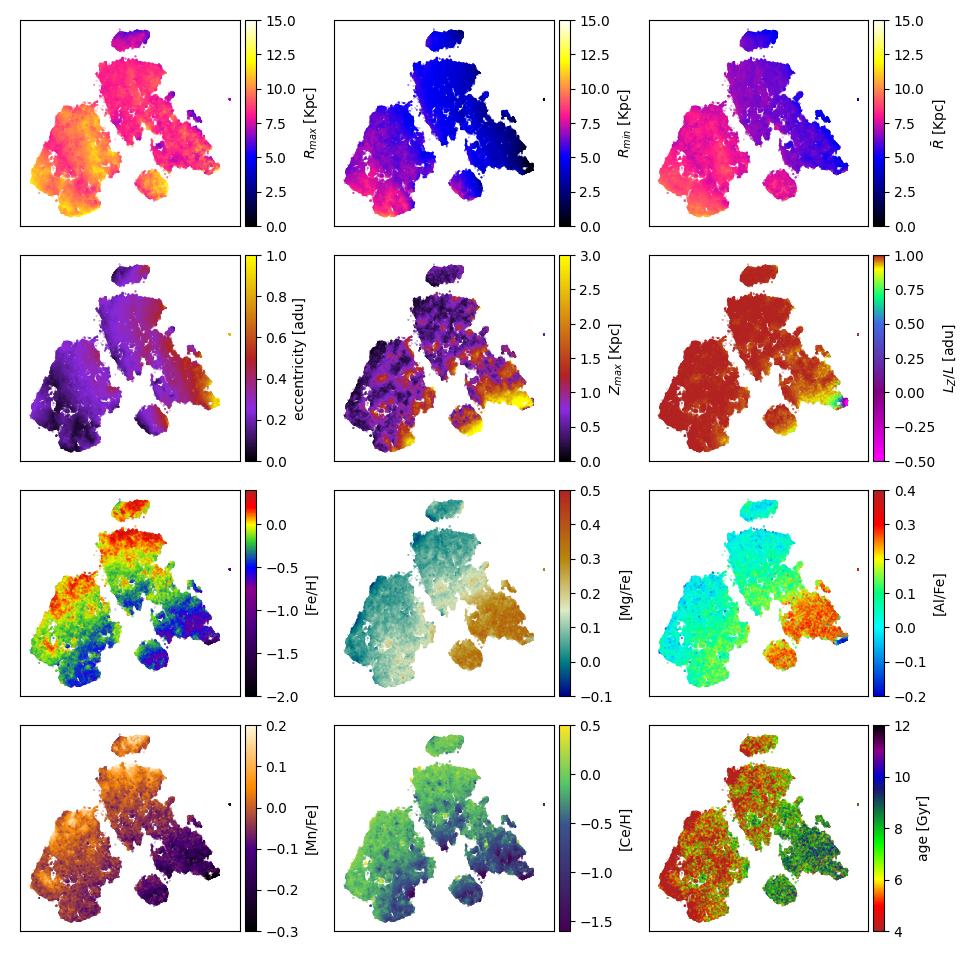}
    \caption{APOGEE manifold colour-coded by the chemical and dynamical parameters selected for the analysis.}
    \label{apogee_variables}
\end{figure*}

\begin{figure*}
    \centering
    \includegraphics[width=\linewidth]{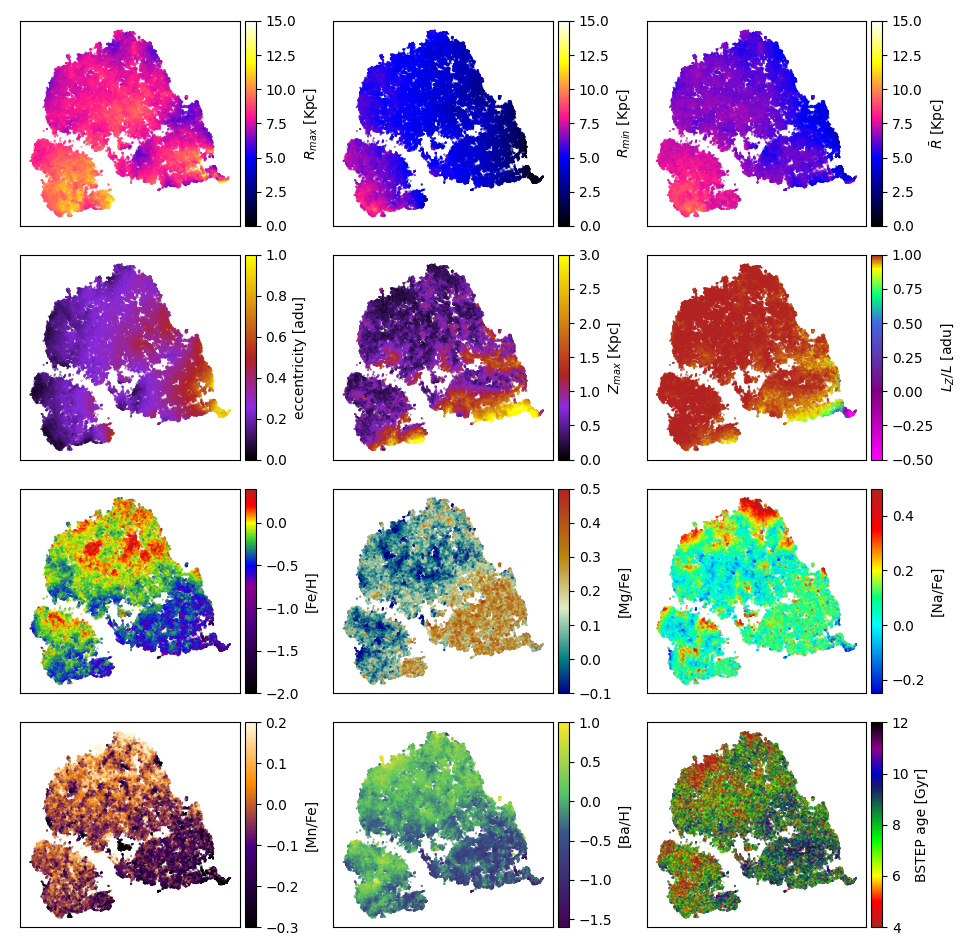}
    \caption{GALAH manifold colour-coded by the chemical and dynamical parameters selected for the analysis.}
    \label{galah_variables}
\end{figure*}

The parameter-coloured manifolds are the most important figures that enable the distinction of characteristic groups by a careful analysis and comparison of each panel, where the colour maps of the variables were chosen to highlight different ranges of values and bring out features on the manifolds to segregate the populations.
These are shown in Figure \ref{apogee_variables} for APOGEE and Figure \ref{galah_variables} for GALAH. 

\begin{figure*}
    \centering
    \includegraphics[width=0.495\linewidth]{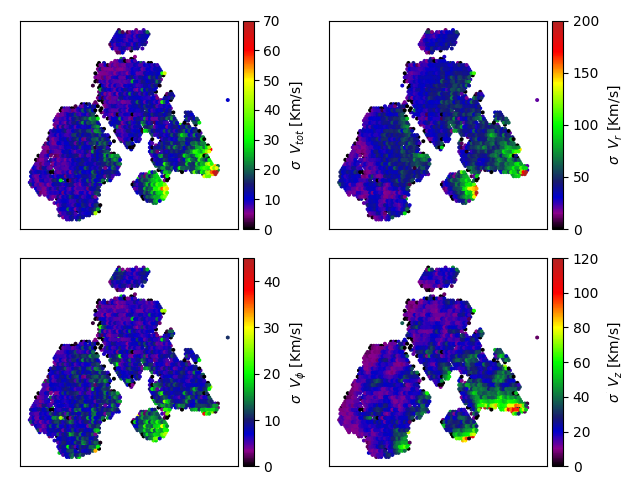}
    \includegraphics[width=0.495\linewidth]{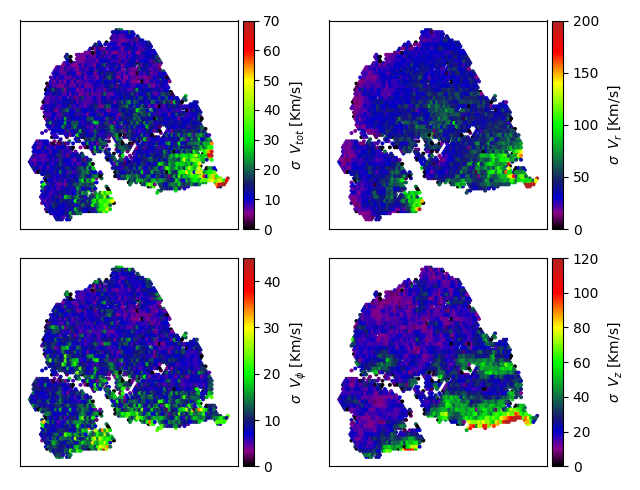}
    \caption{APOGEE (left panel) and GALAH (right panel) manifolds colour-coded by total velocity dispersion and its radial ($V_{r}$), tangential ($V_{\Phi}$) and disk plane transversal ($V_{Z}$) components. }
    \label{vel_dispersion}
\end{figure*}

Also, an important measure for the kinematics of the populations is the velocity dispersion in the $V_{r}$, $V_{\Phi}$ and $V_{Z}$ components, and this was made by binning the manifold into hexagons and taking the standard deviation of the velocities inside each bin. The bin size was chosen in a way that structures were preserved while keeping at least 5 stars into each bin.
The velocity dispersion manifolds with hexagonal binning are shown in Figure \ref{vel_dispersion} for APOGEE and GALAH.
It is possible to verify the match between populations with higher eccentricities and larger $\sigma V_{r}$ in the manifolds, and also the match of those with higher $Z_{max}$ and larger $\sigma V_{z}$ as a good sanity check for the kinematics consistency.

The selected groups were drawn by hand, carefully inspecting the parameter-coloured and the velocity dispersion manifolds to produce consistent populations for both surveys at the same time, searching for gaps in the clusters and leaps in the values of the colour-coded variables that could delineate the region boundaries.
This was possible due to the final configuration of the manifolds, showing the main Galactic features roughly on the same locations for both APOGEE and GALAH, which would be extremely difficult to achieve in other implementations of t-SNE, where the fine-tuning of the convergence steps plays a major role in the desired result.

By analysing the Figures \ref{apogee_variables} and \ref{galah_variables}, the main contrast between the survey samples is the scattering and visual separability.
Although the orbital parameters are derived using the same source, the chemical abundances have a much larger spread in GALAH than APOGEE, which translates into more concentrated and distinct regions for the latter. 
The ages, as stated before, were not used as t-SNE input features, but are valuable markers for these older populations and in-survey comparison.
Many of the labelled groups carry an inner- or outer- designation that refers mainly to their $R_{avg}$ location, appearing in contrast at the upper-right panels of Figures \ref{apogee_variables} and \ref{galah_variables}, and further in Figure \ref{apogee-ravg} as a bimodal distribution of the whole sample of APOGEE.
In general, "inner" groups are for $R_{avg} < 7$ kpc and "outer" are for $R_{avg} > 7$ kpc.

Nevertheless, it is easy to identify the thick disk due to its alpha-enhanced character, the lower overall metallicity, higher $Z_{max}$ and tilted orbits in relation to the Galactic plane (lower $L_z/L$), and especially higher velocity dispersions in the lower-right portion of the manifolds in Figures \ref{apogee_variables}, \ref{galah_variables} and \ref{vel_dispersion}.
The thick disk appears splitted into two components: the \textit{inner thick disk} attached to the corotation, and the \textit{outer thick disk} associated with the $\Bar{R}_{o}$ concentration at R$\sim$8-9 kpc, and this \textit{outer} component has a contrasting high $\sigma V_{\phi}$ in comparison with the rest of the manifold.
In both APOGEE and GALAH it is possible to spot a slightly older concentration, which shows larger radial velocity dispersions and eccentricities, and will be discussed further in Section \ref{thickdsec}.
At the bottomost region of this population we can also notice a dynamically hotter portion with even higher values of $Z_{max}$ and highly tilted orbits, achieving negative values (retrograde orbits) in its right end. 
This portion is the so-called \textit{heated} or \textit{splashed} thick disk that is associated with the merger with GSE, where its right end is a piece of the dynamical halo including the accreted populations with lower Mn, Na and Al abundances.

\begin{figure}
	\includegraphics[width=\columnwidth]{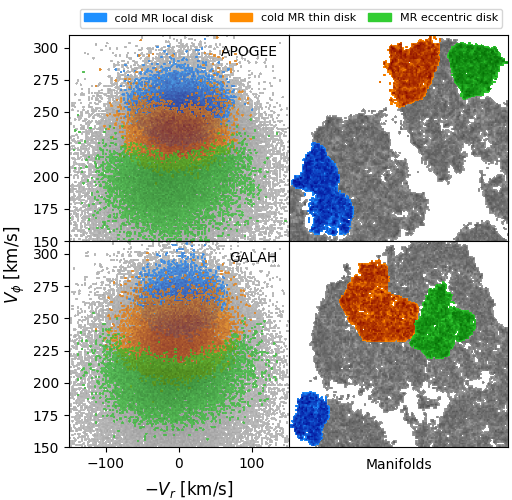}
    \caption{$V_{\phi}-V_{r}$ plane (left panels) and respective manifolds (right panels) for APOGEE and GALAH metal rich populations, showing \textit{metal-rich eccentric disk} (green), \textit{cold metal-rich thin disk} (orange) and \textit{cold metal-rich local disk} (blue) regions.}
    \label{movingroups-manifold}
\end{figure}

The upper-central region is mostly comprised of dynamically colder thin disk stars with higher metallicities, mostly contained inside the solar galactocentric radius within an \textit{inner disk}. 
There are many interesting features in this structure, and one of them is the higher eccentricity portion in the right with higher $V_{\phi}$ dispersion (labelled \textit{metal-rich eccentric disk}), which according to Figure \ref{movingroups-manifold} could be associated to the Hercules moving group, where \cite{PerezVillegas_2017} pointed that this bar-induced dynamical population is composed of stars that move between the solar radius and the innermost regions of the Galaxy, showing itself as an overdensity in the U-V plane (Figure 2 of their paper) peaking roughly at -30 km/s in U and -50 km/s in V, which translates into $(V_{r},V_{\phi})=(18.9, 200.24)$ km/s considering their LSR at (U, V) = (-11.1, -12.24) km/s, and a local galactic rotation of 238 km/s, whereas our \textit{metal-rich eccentric disk} populations peaks in density at $(V_{r},V_{\phi})=(24.34, 200.28)$ km/s for APOGEE and $(V_{r},V_{\phi})=(16.4, 203.63)$ km/s for GALAH.
Also, the labeled \textit{cold metal-rich thin disk} and \textit{cold metal-rich local disk} could be associated with the moving groups of Hyades and Sirius, respectively, according to their locations in the $V_{\phi}-V_{r}$ plane (analogous to U-V plane) as shown by \cite{Bernet2022}.
At the lower end of this portion, we can spot a small low-metallicity tail, being the inner counterpart of the metal-poor populations of the outer portion (bottom-left region).
The proportion of metal-poor stars between inner and outer populations reproduces the metallicity gradient that increases towards the Galactic centre (e.g., \citealt{sun2024mapping}), where the outer portion carries a noticeably larger metal-poor fraction than the inner portion for the thin disk, and no significant radial gradient when considering the \textit{outer thick disk} and \textit{inner thick disk}.

The bottom-left region is mostly composed of an outer thin disk (with the exception of the high alpha blob in the GALAH manifolds, that is associated with the thick disk) reaching the solar Galactic radius and some metal-poor structures, consistent with the negative metallicity gradient towards the outer parts of the Milky Way.
This \textit{outer thick disk} portion, as shown in salmon in Figure \ref{tsne_convergence}, stays associated with the thick disk cluster during the convergence, but after the final annealing with lower perplexities, it tends to associate with the \textit{outer disk} region in GALAH, and as a separated cluster in APOGGE, closer to the bulk of the thick disk.

Also, a \textit{transition zone} appears between the cold, thin, metal-rich disk and the alpha-enhanced thick disk, with moderate chemical features for virtually all of the cases including the ages, where this region is clearly older when compared to the thin disk.
Similar to some of the selected regions, this zone shows a few sub-structures that will be discussed further in Section \ref{transition}.

One remarkable difference between the two surveys is the size proportions of the clusters, which are explained by the sky sampling of each one. APOGEE scans mostly the northern sky, which points towards the Galactic anticentre, and thus there are more outer disk stars.
In the other hand, GALAH scans the southern sky in the direction of the Galactic centre, so more inner (with respect to the solar radius) disk stars are expected in the sample.
Moreover, GALAH basically avoids the Galactic plane (with the exception of the Galactic centre) while APOGEE samples its available sky more uniformly.

\section{Discussion}
\label{discussion}

It is important to remember that, as a technique that minimises the similarity distribution divergence, the t-SNE manifolds can be seen as similarity maps of the Galaxy.
In this context, the relative distances between the selected groups can give hints on how they are related in a Galactic assembly scenario.
In the following topics, we will discuss each of the main components, new insights and some caveats that are intrinsic to the problem.

\subsection{Thin disk}
\label{thindsec}

\begin{figure}
	\includegraphics[width=\columnwidth]{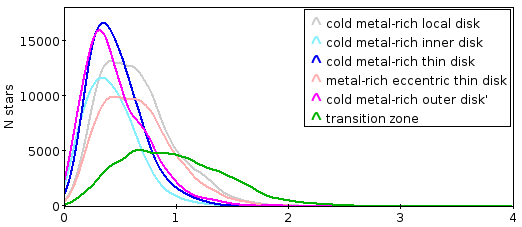}
	\includegraphics[width=\columnwidth]{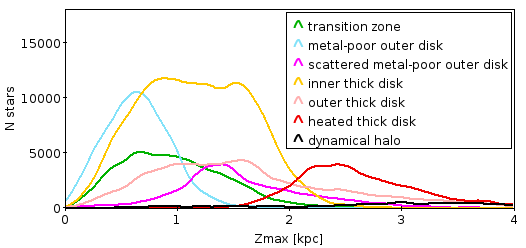}
    \caption{Distribution functions of $Z_{max}$ for the dynamically cold metal-rich (upper panel) and hot metal-poor (lower panel) populations of the APOGEE sample.}
    \label{apogee_zmax}
\end{figure}

First, we need to stress that due to the poor sampling of the Galactic plane in the direction of its centre (see Figure \ref{aitoff_map}), we miss the bulk of thin disk stars confined to the lowermost vertical elongations, but still within a reasonable range considering stellar distances, proper motions and radial velocity determination errors, which can broadly affect the orbital parameters.
Our thin disk sample, due to this, shows some dispersion of $Z_{max}$, but still more concentrated within Z$\sim$1 kpc as shown for a selection of metal-rich portions of the manifolds in Figure \ref{apogee_zmax}, together with thick disk populations for a comparison.
These thin disk populations can be easily spotted through low $Z_{max}$, young ages and their typical [Fe/H] ranging from -0.25 to +0.3 dex on the colour-coded manifolds of Figures \ref{apogee_variables} and \ref{galah_variables}, as well as the low values of velocity dispersion in Figure \ref{vel_dispersion}.
We show mostly APOGEE in the figures of this section, mainly for its larger sampling of the outer disk.

\begin{figure}
	\includegraphics[width=\columnwidth]{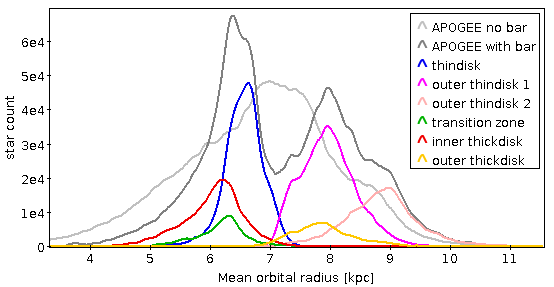}
    \caption{Distribution functions of mean orbital radius for the APOGEE sample. In light gray we show the total sample for the orbits calculated with the axisymmetric potential, and in dark gray for the barred potential used in this study. Blue line shows the central manifold portion (\textit{inner thin disk}).  \textit{Inner thick disk} (red), \textit{outer thick disk} (yellow) and \textit{transition zone} are shown for comparison.}
    \label{apogee-ravg}
\end{figure}

One of the main differences between the two major regions containing the thin disk is their mean orbital radius $\Bar{R}_{o}$, where the central manifold portion is bound to a concentration centered at R$\sim$6.4 kpc with a long tail towards the Galactic centre, and the lower-left portion is associated to another concentration distributed between R$\sim$8-9 kpc, as illustrated in Figure \ref{apogee-ravg} for the case of APOGEE, which better covers the mean orbital radius space.
The outer disk was split into the more metal-rich upper portion (\textit{outer disk 1} in magenta, including the \textit{cold metal-rich local disk} and \textit{cold metal-rich outer disk}) and a more metal-poor with larger $R_{min}$ at the bottom portion (\textit{outer disk 2} in salmon, including the \textit{metal-poor outer disk} and \textit{scattered metal-poor outer disk}) to show the origin of its bimodality.
We show the mean orbital radii of the total APOGEE sample calculated with the barred galactic potential aforementioned in comparison with the MW2014 axisymmetric potential (also with twice the original halo mass).

Our assumed Galactic potential for orbital integration with a bar pattern speed of $\Omega_{b}$=39 km/s/kpc places the corotation radius around $\sim$6 kpc, but the locus of the resulting radial density caused by the corotation is subject to other parameters and effects such as bar length and the aforementioned phase space measurement accuracy, therefore we associate the R$\sim$6.4 kpc overdensity as the corotation effect on mean orbital radius of the sample stars. 
As early stated by \cite{athanassoula1980}, growing galactic bars can drive the spiral arm structure, and \cite{garma2021} have shown through simulations that the spiral arm pattern is the collective result of the rotating bar perturbing the stellar orbits, thus forming the arms through a density-wave mechanism.
\cite{dias2019} studied the spiral arm structure and placed its corotation radius at R=8.5 kpc, which is tightly reproduced in our mean orbital radius density distribution (where the \textit{outer} populations lie), leading us to suggest that the Gaia observed phase space combined with the Galactic potential used here could be able to reproduce by itself the spiral arm structure of the Galactic disk.

\begin{figure}
	\includegraphics[width=\columnwidth]{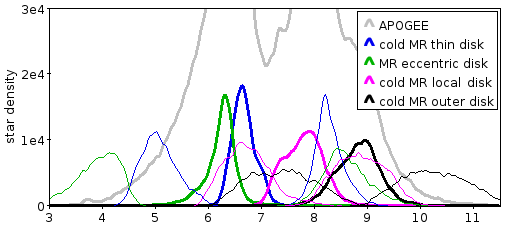}
	\includegraphics[width=\columnwidth]{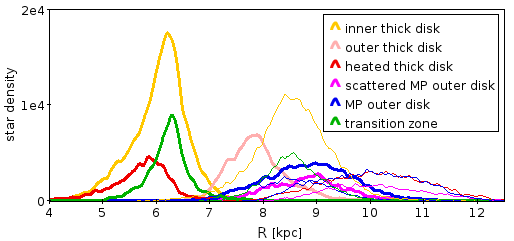}
    \caption{Radial distributions for the metal-rich (upper panel) and metal-poor (lower panel) populations of the APOGEE sample with the mean orbital radius plotted in thick lines. The thin lines represent the apocentric and pericentric radii of the metal-rich populations, and the apocentric radii of the metal-poor populations.}
    \label{apogee_raprper}
\end{figure}

Moreover, the density waves are the main star formation locations, as their potential favours converging gas inflows \citep{pettitt2019}.
The dynamically cold metal-rich populations show this mean orbital radius binding to the corotation of the bar and the spiral arms, and when we inspect the apocentric and pericentric radius of these populations (upper panel of Figure \ref{apogee_raprper}), it is possible to notice that most of their extreme orbital radii are concentrated in the bar and spiral arm corotations.
\cite{Lucey_2023} constrains a dynamical bar length (maximum extent of trapped orbits) of 3.5 kpc with an overdensity spanning up to 4.8 kpc, representing an upper limit for the direct bar churning influence.
The \textit{metal-rich eccentric disk}, associated with the Hercules moving group, dives more deeply inside the the churning influence zone of the bar, and consequently the inner Galaxy, which might explain its slightly older right-end in the age manifold of Figure \ref{apogee_variables}.
Moreover, by closely inspecting the orbits that belong to Hercules in Figure 3 of \cite{PerezVillegas_2017} ((d), (g), (h) and (i) panels), their apocentric and pericentric radii seem to coincide with the ones at Figure \ref{apogee_raprper}, providing more evidences that the \textit{metal-rich eccentric disk} is the isolated Hercules stream.
The \textit{cold metal-rich local disk} population, where the sun lies, even if associated with the \textit{outer} portion, has a lower apocentric radius coincident with the spiral arm corotation radius, like most of the metal-rich populations, feature also visible in the GALAH manifolds on Figure \ref{galah_variables}.

\begin{figure}
	\includegraphics[width=\columnwidth]{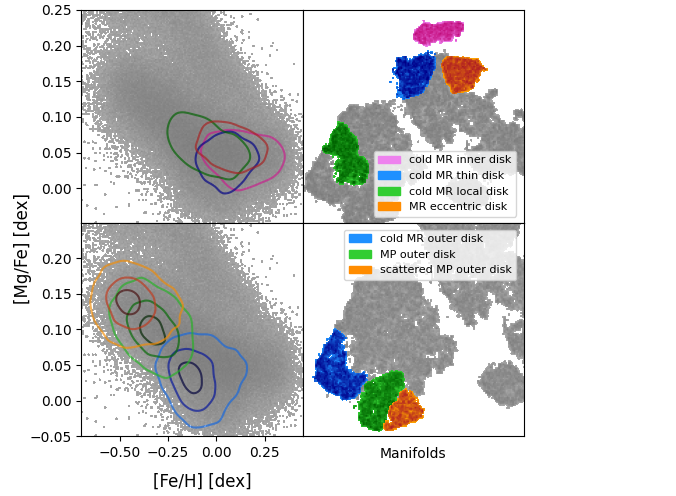}
    \caption{Tinsley-Wallerstein diagrams (left panels) with contours representing the 50th percentile of the metal-rich sequence(upper left), 16th, 50th and 84th percentiles of metal-poor sequence (lower left), and respective manifold locations (right) for selected thin disk populations in APOGEE. \textit{metal-rich} and \textit{metal-poor} are abbreviated to MR and MP, respectively.}
    \label{apogee_mgfe}
\end{figure}

Digging more deeply into the chemical behaviour of these populations, we show in Figure \ref{apogee_mgfe} the locations in the Tinsley-Wallerstein diagram of the selected inner and outer populations for APOGEE (due to the better [Mg/Fe] resolution), where the outermost ones lie along a metal-poor sequence on the left portion of the lower [Mg/Fe], and the innermost ones populate its metal-rich end, commonly associated as the younger population.
A further discussion of the inner metal-rich population is presented in Section \ref{mrpops}.

\begin{figure}
	\includegraphics[width=\columnwidth]{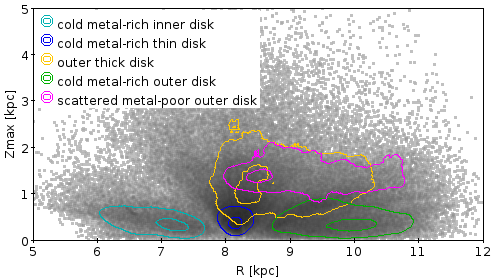}
	\includegraphics[width=\columnwidth]{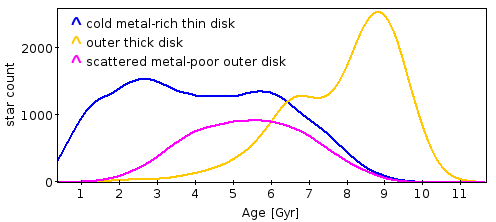}
    \caption{$Z_{max}$ vs R plane for APOGEE sample (grey scattered points) with contour plots of the \textit{scattered metal-poor outer disk} compared with other selected components in the in upper panel, and age distributions in the lower panel.}
    \label{outer_mp_scat}
\end{figure}

The \textit{scattered metal-poor outer disk} shows an alpha enhancement and larger velocity dispersions, and in Figure \ref{outer_mp_scat} we show the behaviour of this population in APOGEE (as it samples more of the outer disk) in function of galactocentric radius and $Z_{max}$ compared to the other thin disk components. 
\cite{das2024} presented results of the heating effect caused in the outer disk by the interaction of a satellite galaxy (probably Sagittarius dSph), and in their Figure 4 a clear bump in $\sigma_{z}$ is visible between $R\sim9-10$ kpc, comprehended in the same place where the \textit{scattered metal-poor outer disk} is located, showing a behaviour more related to the thick disk in comparison with the other thin disk selected populations.
Besides the location, the age distributions at the lower panel of Figure \ref{outer_mp_scat} shows the \textit{scattered metal-poor outer disk} near $\sim$6 Gyr, and also the [Fe/H]$\sim$-0.5 together with [$\alpha$/Fe]$\sim$+0.14 (see Figure \ref{apogee_mgfe}) are in agreement with \cite{das2024} results.
This suggests that this population may have a connection to the passage of Sgr dSph.
In this panel we also show populations belonging to thin and thick disk for comparison.
In Section \ref{transition}, we give some insights about the transition zone and its possible relation to this disk perturbation event.

\subsection{Inner metal-rich population}
\label{mrpops}

\begin{figure}
	\includegraphics[width=\columnwidth]{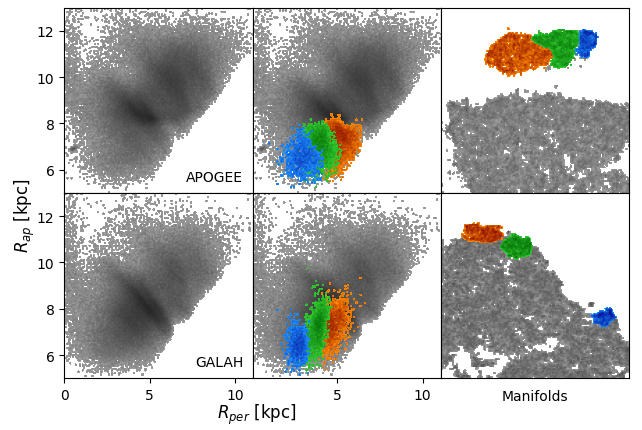}
    \caption{$R_{ap}$-$R_{per}$ plane for APOGEE and GALAH, with details of the \textit{cold metal-rich inner disk} cluster sub-divided into 3 pieces. The blue substructure is an innermost and more eccentric portion of the region that appears apart from the other 2 in the GALAH manifold.}
    \label{coldinnermr}
\end{figure}

As stated before, we kept the kinematic homogeneity by using only Gaia DR3 measurements for orbital constraints, but some artefacts could arise from the use of different chemical data when comparing the manifold results, and also from the different spatial sampling of APOGEE and GALAH.
As an example, the assigned \textit{cold metal-rich inner disk} appears as an isolated island in APOGEE, but for GALAH, the definition of an analogous structure is not trivial, since the sampling of the surveys is not uniform in a specific region of the $R_{ap}$-$R_{per}$ plane as shown in Figure \ref{coldinnermr}.
Moreover, the light element used for GALAH was Na, while for APOGEE it was Al, making it difficult to make a direct comparison for this specific metal-rich and Na-rich population.
This \textit{cold metal-rich inner disk} structure has a lower $R_{ap}$, confined to the inner parts of the Galaxy, and was found to be splitted on the GALAH manifold in 3 sub-structures, also shown in Figure \ref{coldinnermr}.
These 3 sub-structures show a reasonable coincidence in the $R_{ap}$-$R_{per}$ plane with the APOGEE counterpart, but it is difficult to tell if they represent the exact same elements.
Moreover, still stressing the incompatibility caveats between surveys, this inner-metal-rich population does not appear alpha-enhanced in APOGEE, and one possible reason is the lower sampling of the Galactic central directions in comparison with GALAH.

\begin{figure}
	\includegraphics[width=\columnwidth]{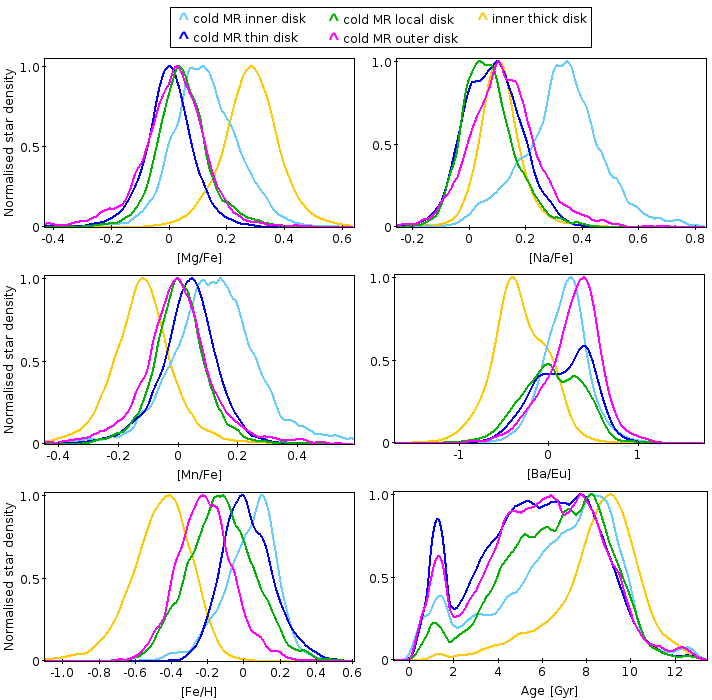}
    \caption{Abundance and age distribution functions of GALAH for the metal-rich populations ([Fe/H] $\gtrsim$ -0.2) \textit{cold metal-rich inner disk}, \textit{cold metal-rich thin disk} and \textit{cold metal-rich outer disk} structures, together with a representative of the thick disk and the local disk for comparison. \textit{Metal-rich} is abbreviated to MR in the legend.}
    \label{coldinnermrDF}
\end{figure}

An interesting feature of this selected portion is the high Na and Mn together with moderate Mg content (Figure \ref{galah_variables}), where its distribution function is centered at $\sim$0.1 dex, with enhanced Mn and Ba/Eu shown in Figure \ref{coldinnermrDF}.

The high Mn and moderate Mg points towards a fast enrichment in an already metal-rich environment, since the yield of Na starts to increase above 2 solar masses \citep{smiljanic2016}, but the higher [Ba/Eu] indicates that the s-process operated significantly in the enrichment of this population.
This same behaviour of [Ba/Eu] is observed in the \textit{cold metal-rich thin disk} and even stronger in the \textit{cold metal-rich outer disk} structures, but with lower Na, Mg and Mn distributions.
\cite{karakas2018} shows that the most significant production of the heavy s-process elements takes place in stars from $\sim$1.15 to 4 solar masses, which covers an evolutionary timescale range between $\sim$0.3 to 7 Gyr.
This result favours the aforementioned secular enrichment of Ba, taking into account the Galactic scenario, which for inner parts of the Galaxy the more active star formation rapidly recycles the Ba made available by the constantly evolving stars, while in the outer portions, the lower star formation rate and thus lower r-process supernovae yields raises the [Ba/Eu] ratio at a lower pace.

If we consider that the GSE merger took place between 10-8 Gyr ago, a large infall of gas should have taken place \citep{deason2013}, and if we take the peak age of the thick disk (around $\sim$9 Gyr in the last panel of Figure \ref{coldinnermrDF}) as a proxy to the merger event, the peak distance between the thick disk and the \textit{cold metal-rich inner disk} is roughly $\sim$1 Gyr.
The peak abundances of Na for the \textit{cold metal-rich inner disk} around $\sim$0.4 dex translates roughly into $\sim$3 solar masses of progenitor stars via the relation presented in \cite{smiljanic2016}, representing a stellar lifetime around $\sim$650 Myr, which is reasonable enough to consider the fast recycling scenario that led to the observed Na and Ba abundances of this population.
Wrapping these pieces together, it is not preposterous to suggest that this population may be the direct cause of the GSE merger reflected onto the inner disk of the Galaxy.
Of course, the BSTEP stellar age estimations have large associated errors of around $\sim$3 Gyr for these populations, but working with the age distributions and its offsets are the best tools available to this kind of analysis.

As for the \textit{cold metal-rich local disk}, encompassing the solar neighbourhood, the [Ba/Eu] ratio shows a broad distribution around the solar value, perhaps signalling that the sun lies in a region where the Ba recycling is not so intense and the Eu production rates are enough to settle down the [Ba/Eu] ratio.

\subsection{Thick disk}
\label{thickdsec}

Until now, most studies worked by dividing 2D diagrams (eg. Tinsley-Wallerstein, Toomre, etc) to distinguish between thin and thick disk and conduct their analyses, but of course, some contamination is expected from this methodologies.
Here we have used a selection of kinematic and chemical properties to segregate these older components with a clearer boundary, being able to visualise at which limit we admit stars to a specific sampling by making use of the colour-coded manifolds of Figures \ref{apogee_variables} and \ref{galah_variables}.

Still from the APOGEE sample (due to its better [Mg/Fe] separation), we trace the $Z_{max}$ distributions of the metal-poor populations in the lower panel of Figure \ref{apogee_zmax}, evidencing the dynamically hot character of the canonical thick disk, and the evident presence of the \textit{heated} (or \textit{splashed}) thick disk with a long tail towards larger Z. 

We can also see the \textit{scattered metal-poor outer disk} exhibiting larger Z elongations, where the mean orbital radius and apocentric radius of the metal-poor populations are shown in the lower panel of Figure \ref{apogee_raprper}. 
The maximum Z elongation is achieved near the apocentre of the orbit, which can be up to R$\sim$12 kpc for this local volume sampling.
The mean orbital radius and apocentric accumulation near the corotation radii also rise for the thick disk, indicating that the bar and spiral arm structure also affects the dynamically hot populations.

\begin{figure}
    \includegraphics[width=\columnwidth]{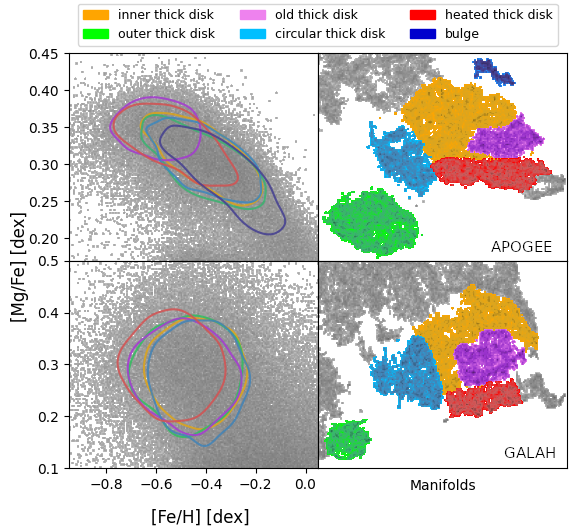}
    \caption{Tinsley-Wallerstein diagrams (left panels) with contours representing the 50th percentile, and manifolds (right panels) for selected thick disk populations of APOGEE (upper panels) and GALAH (lower panels). The \textit{circular thick disk} and the \textit{old thick disk} are superposed to the \textit{inner thick disk}.}
    \label{mgfe_dist_manifolds}
\end{figure}

The so-called \textit{heated thick disk} was interpreted by \cite{DiMatteo_2019} as the imprint left by a great merging event (GSE) heating the old Galactic disk.
Here we have isolated this population that is significantly hotter and with much larger velocity dispersions than the canonical thick disk, as shown in Figures \ref{vel_dispersion} and \ref{apogee_zmax}.
This population can be distinguished not only by its kinematics, but also for the chemical properties, where its [Mg/Fe] distribution is clearly shifted towards higher values, and for [Fe/H], shifted towards lower values (shown in Figure \ref{mgfe_dist_manifolds}), which corroborates with the proposition that this \textit{heated thick disk} was an older disk being splashed by a merger event.
Here we can see the similarity space brought into action when it comes to the relations between populations, showing the \textit{heated thick disk} as a bridge between the \textit{thick disk} and the \textit{dynamical halo}.

By a closer inspection of the colour-coded manifolds in the thick disk region (also in Figure \ref{mgfe_dist_manifolds}), we could identify 2 sub-structures of the \textit{inner thick disk}, which are the \textit{circular thick disk} with low eccentricities, and the \textit{old thick disk} with slightly older ages.

The separability of the populations is clearer for APOGEE, and we can see that the \textit{old thick disk} concentrates at the most metal-poor and alpha-rich end, followed by the \textit{heated thick disk}.
The \textit{old thick disk} together with the more general \textit{inner thick disk} and \textit{outer thick disk} occupy roughly the same distribution along the alpha-rich population, while the \textit{bulge} extends up to the more metal-rich end as expected for the inner Galaxy.

Apparently, the more circular orbits are the only feature that distinguish this \textit{circular thick disk} population, and we did not find any other unique chemical or kinematic signs other than the apocentric and pericentric radii coinciding with the \textit{cold metal-rich thin disk} population and the [Mg/Fe] and [Fe/H] distributions matching those of the \textit{outer thick disk}.

\begin{figure}
	\includegraphics[width=\columnwidth]{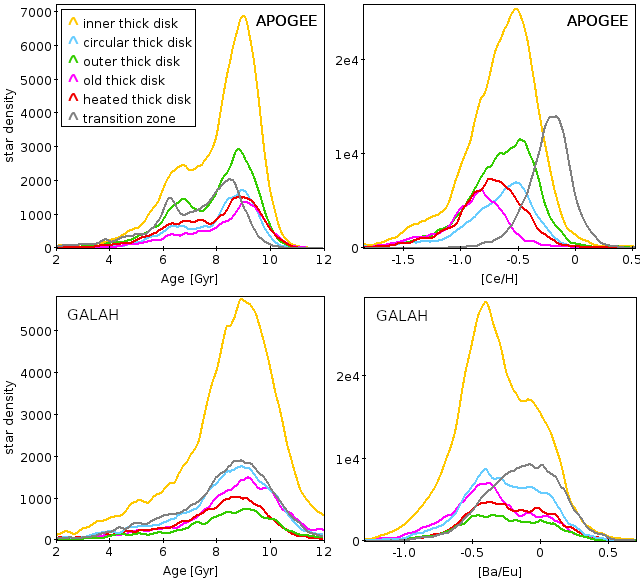}
    \caption{Age (left panels) and heavy elements (right panels) distributions for the thick disk sub-populations in the GALAH (lower panels) and APOGEE (upper panels). The bump at $\sim$6 Gyr on APOGEE's astroNN appears as a slight asymmetry in the GALAH distributions, and the bimodality can be seen in the [Ba/Eu] distribution.}
    \label{age_distributions}
\end{figure}

As for the \textit{old thick disk}, it can be spotted as a lower age region in the heart of the thick disk portion of the parameter-coloured manifolds (Figures \ref{apogee_variables} and \ref{galah_variables}), and we find the [Mg/Fe] and [Fe/H] distributions in Figure \ref{mgfe_dist_manifolds} even more shifted than the \textit{heated thick disk}.
Moreover, if we inspect the age distributions of the thick disk (Figure \ref{age_distributions}), this population appears slightly, but not negligibly shifted towards older ages for both survey samples, effect that is also observed in the [s-process/H], as proposed here as a possible chemical clock.
This population could be either just an "older tail" of the thick disk or a fossil record of a primordial Galactic configuration, where we favour this scenario in the following section.

A small portion of the outermost extents of the bulge population also appears, but only clear and isolated in the APOGEE. 
Our GALAH selection even samples the central Galactic region, but the bulge may be too attached to the thick disk to reliably isolate it. 
The \textit{bulge} population can be spotted as a stretched clump above the thick disk region in the APOGEE manifold, and as shown in Figure \ref{apogee_variables}, it is among the lowest apocentric and pericentric radii, and share the same chemical pattern as the thick disk, with Z$_{max}$ contained inside $\sim$1.5 kpc.

\subsection{Transition zone}
\label{transition}

\begin{figure}
	\includegraphics[width=\columnwidth]{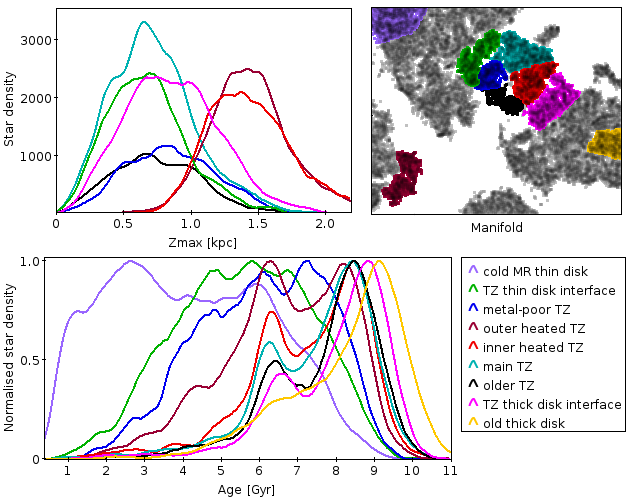}
    \caption{Transition zone splitting showing Z$_{max}$ distributions in the upper left panel, manifold with the selected regions in the upper right panel, age distributions in the lower left panel, and colour-coded regions in the lower right legend, with \textit{transition zone} abbreviated to TZ and \textit{metal-rich} to MR. For comparison, we also show the \textit{cold metal-rich thin disk} and the \textit{old thick disk} in the extremes.}
    \label{transition_split}
\end{figure}

The so-called \textit{transition zone} is an intermediate population with properties that stay between the thin and thick disks, and was interpreted by \cite{Recio_Blanco_2014} as a settling process that the thick disk experienced, increasing its rotation and lowering the tangential velocity dispersion, what can be indeed observed in Figure \ref{vel_dispersion}.
We will refer to the APOGEE sample for this analysis, since the transition zone appears more clear and defined in the manifolds.
The $Z_{max}$ distribution is compared to other populations in Figure \ref{apogee_zmax}, and follows a pattern that stands between the metal-poor populations and the thick disk, showing a longer tail towards larger values.
Its chemical and age patterns are rather puzzling, as can be explored in Figure \ref{apogee_variables}, showing a moderate metallicity around [Fe/H]$\sim$-0.25, moderate [$\alpha$/Fe]$\sim$0.15, but enriched [Ce/Fe] (comparable to the metal-rich populations) and older ages (comparable to the thin disk) as seen in the manifolds.
We have splitted the transition zone and its surroundings in the manifold to better dissect its properties, shown in Figure \ref{transition_split}. 

Besides its main portion (shown in yellow at Figure \ref{final_manifolds}, connecting the \textit{inner} thin disk and the bulk of the thick disk), the \textit{outer} portion of the manifold also has a counterpart (labelled \textit{outer heated transition zone}) that shares some characteristics of the transition zone, already disconnected from the \textit{outer thick disk} blob.
By inspecting the APOGEE convergence map of Figure \ref{tsne_convergence}, we can see in the 100$^{th}$ iteration of \textit{annealing 3} (lowest rightmost panel) that the \textit{outer heated transition zone} and the \textit{outer thick disk} were once connected, representing their similarity relation which can be seen as the \textit{outer} counterpart of the transition zone.
Moreover, the aforementioned \textit{scattered metal-poor outer disk} appears also very close and even stretched upon the \textit{outer thick disk}, which also favours its connection with this \textit{transition zone}.

In the Z$_{max}$ distributions of Figure \ref{transition_split}, the "heated" character of the \textit{inner heated transition zone} and \textit{outer heated transition zone} becomes clear and contrasts with the remaining populations of the \textit{transition zone}.
These two populations also have the strongest bimodality among the age distributions, whose peaks are located at $\sim$6 Gyr and $\sim$8-9 Gyr, which coincides with the Saggitarius dSph disturbance \citep{das2024} and the GSE merger, respectively.
The \textit{outer heated transition zone} shows the largest ratio of the 6 Gyr peak compared to the 8 Gyr peak, corroborating with \cite{das2024} results of a heating effect in the outer regions of the disk.
In the other hand, this ratio gradually decreases for \textit{inner heated transition zone}, \textit{main transition zone}, \textit{old transition zone}, until it reaches the \textit{thick disk interface} and a very weak character in the \textit{old thick disk}.

\begin{figure}
	\includegraphics[width=\columnwidth]{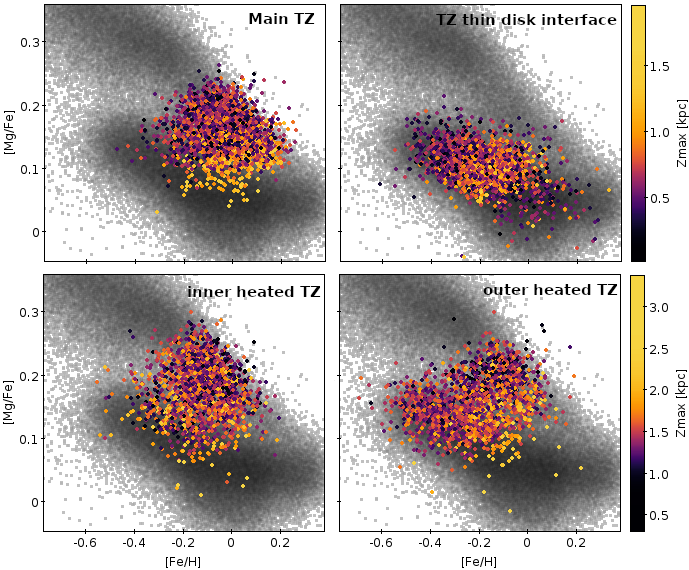}
    \caption{Tinsley-Wallerstein diagrams with \textit{transition zone} populations colour-coded by Z$_{max}$, showing its gradients for the case of APOGEE.}
    \label{apogee_mgfe_trans}
\end{figure}

If we inspect the Z$_{max}$ behaviour of some of these populations in the Tinsley-Wallerstein diagram of Figure \ref{apogee_mgfe_trans}, an even more puzzling effect appears: for all the cases shown, the Z$_{max}$ value increases for decreasing [Mg/Fe], the opposite of what would be expected in the evolutionary sense.

This effect partly contradicts the settling process proposed by \cite{Recio_Blanco_2014}, which leads to a scenario where low-alpha thin-disk stars were scattered to higher-Z$_{max}$ orbits, and the older, dynamically colder stars in the moderate-alpha sequence are the ones that were probably formed in a post-merger environment (after GSE), as seen in the age distributions of Figure \ref{transition_split}.
The \textit{metal-poor transition zone} may give some hints about this formation scenario, and was named mainly due to its bluer hue in the [Fe/H] panel of Figure \ref{apogee_variables}, in contrast to the rest of the \textit{transition zone}. 
While its age profile seems more similar to the \textit{thin disk interface}, its tri-modal structure has peaks at ages that coincide with would be an "aftermath" of each disk perturbation, possibly due to metal-poor gas inflow.

\begin{figure}
	\includegraphics[width=\columnwidth]{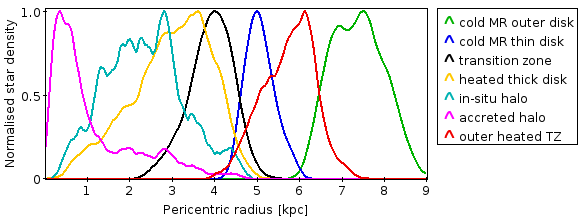}
    \caption{Pericentric radius distribution functions of dynamically cold and dynamically hot populations, with \textit{transition zone} abbreviated to TZ and \textit{metal-rich} to MR.}
    \label{apogee_inner-trans}
\end{figure}

To better understand the Z$_{max}$ heating mechanism, we inspected the pericentric radii of the transition zone in Figure \ref{apogee_inner-trans} together with dynamically hot and cold populations.

The \textit{outer heated transition zone} avoids the stronger churning effects (cited previously in Section \ref{thindsec}) with pericentric radii located mostly at $\sim$6 kpc, which could explain the preservation of the $\sim$6 Gyr peak ratio.
In general, it seems mostly bound to the bar corotation for pericentric radii, spiral arm corotation for mean orbital radius, and outer Lindblad resonance at $\sim$10.5 kpc for apocentric radii.
Moreover, \cite{das2024} shows chemical and age distributions with a bump at [Fe/H]= -0.5, [$\alpha$/Fe] = 0.1 and age = 7.2 Gyr, where \textit{outer heated transition zone} has one of the age peaks exactly at 7.2 Gyr and [$\alpha$/Fe] = 0.1, but at [Fe/H] = -0.3.

Putting these pieces together, we suggest that the transition zone could have a formation history connected to the disk perturbation from satellite mergers followed by a subsequent starburst from the infall of metal-poor gas.

\subsection{Dynamical halo}
\begin{figure}
	\includegraphics[width=\columnwidth]{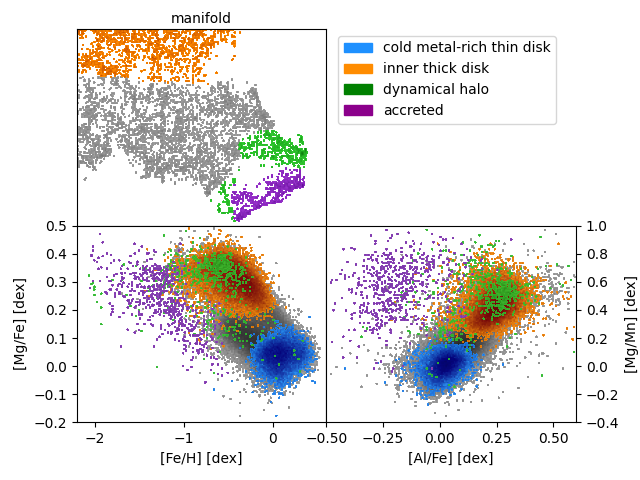}
    \caption{Manifold (upper left), Tinsley-Wallerstein (lower left) and [Mg/Mn] vs [Al/Fe] (lower right) diagrams for the dynamical halo and accreted populations for APOGEE. The accreted population is superposed to the \textit{dynamical halo}. Thin and thick disk representatives are also plotted for comparison.}
    \label{apogee_dyn_halo}
\end{figure}

At the tip of the rightmost end of the \textit{heated thick disk} structures in the manifolds shown in Figure \ref{final_manifolds}, there are stars with highly tilted orbits, together with very low [Fe/H] abundances.
This portion represents the stars belonging to the \textit{dynamical halo} of the Galaxy, where they approach the solar neighbourhood during their dynamically hot orbits, containing populations with high eccentricities, retrograde motions and large velocity dispersions.
For APOGEE, the piece with very low [Al/Fe] abundances represents the accreted population, as shown by \cite{Das2020}, where part of it was interpreted by \cite{Haywood_2018}, \cite{belokurov2018} and \cite{helmi2018} as the remnant of a major merger with a dwarf galaxy between 8 and 10 Gyr ago.
In Figure \ref{apogee_dyn_halo} we present the manifold locations and chemical diagrams for the dynamical halo and the accreted populations, represented together with the other main Galactic components.

\begin{figure}
	\includegraphics[width=\columnwidth]{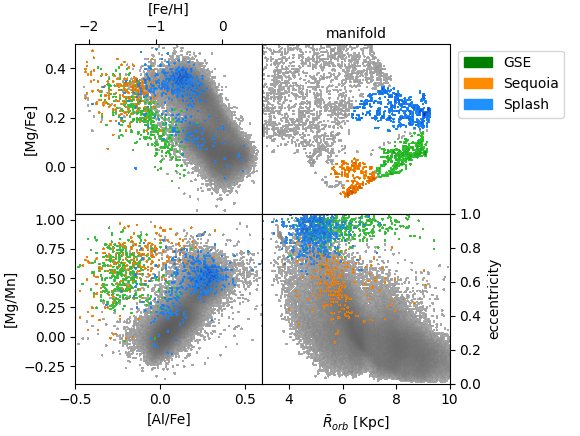}
    \caption{Manifold (upper right), Tinsley-Wallerstein diagram (upper left), [Mg/Mn] vs [Al/Fe] plane (lower left) and eccentricity vs mean orbital radius (lower right) for the segregated dynamical halo populations.}
    \label{dynhalo_pops}
\end{figure}

If we continue to further split this dynamical population in the manifold, it is possible to identify more  structures such as \textit{Splash} (e.g., \citealt{Myeong_2022}) and \textit{Sequoia} (e.g., \citealt{Feuillet_2021}), as shown in Figure \ref{dynhalo_pops}.

For the case of \textit{Sequoia}, it is typically associated with retrograde orbits, but the chemical pattern closely represents the one found by \cite{Feuillet_2021} and the moderate eccentricities reported by \cite{da_Silva_2023}, which is another good example of multidimensional segregation of halo populations using t-SNE.
We have checked the $L_{z}$ behaviour of our \textit{Sequoia}-assigned population with orbits integrated with axisymmetric potentials, and they show the same patterns presented in Figure \ref{dynhalo_pops}, so the bar potential cannot be the culprit.
Also, our restrictive selection could have eliminated the retrograde portion, as streams are sensitive to the spatial sampling.
It is possible that \textit{Sequoia} is not exclusively a retrograde population, as result of the selection cuts of previous studies, but instead distributed along a broad range of positive and negative $L_{z}$.

It is clear that the manifold clumps represent the real structures that were identified in the previous works by using bivariate chemodynamical planes, but without relying on sharp cuts and several planes together to further filter the sample.

\subsection{Open and globular clusters}

\begin{figure}[b!]
	\includegraphics[width=\columnwidth]{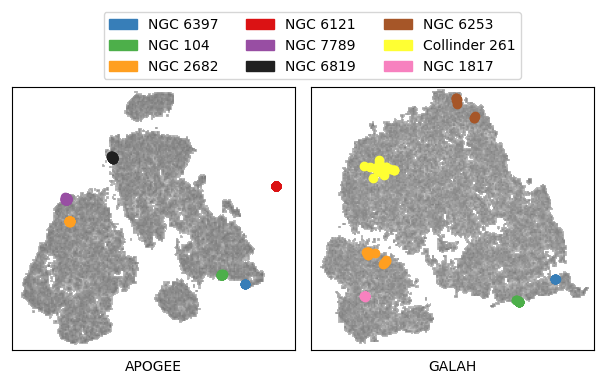}
    \caption{Manifolds of APOGEE and GALAH with the open and globular clusters of each sample.}
    \label{manifold_clusters}
\end{figure}

Both APOGEE and GALAH samples contain stars belonging to clusters, where the open clusters are located together with the \textit{thin disk} population, while the globular clusters stand within the \textit{thick disk} region, or even isolated, as for the case of NGC 6121 (Messier 4).
In Figure \ref{manifold_clusters} we show the cluster locations in the manifolds for APOGEE and GALAH.

The common globular clusters NGC 6397 and NGC 104 appear on the same regions of the manifolds, which are the two extremes of the \textit{heated thick disk}, while the common open cluster NGC 2682 appears at the upper portion of the \textit{outer disk}.
The other open clusters NGC 6819, NGC 7789, NGC 6253, NGC 1817 and Collinder 261 are also located in the cold disk portions.

NGC 6121 appears isolated, and in the colour-coded manifolds of Figure \ref{apogee_variables}, it is possible to see its the yellowish hue, indicating a high eccentricity, with apocentric radii up to 7 kpc and pericentric radii approaching the Galactic center at $\sim$0.5 kpc.
\cite{nordlander_2024} measured new abundances for this cluster, with [Fe/H] = -1.13$\pm$0.07 and [Mg/Fe] = +0.35$\pm$0.06 in agreement within uncertainties with our APOGEE-based values of [Fe/H] = -1.06$\pm$0.05 and [Mg/Fe] = +0.34$\pm$0.05.
\cite{song_2018} reported an isochrone-based age between 12 and 13.3 Gyr for NGC 6121, characterising it as a very old cluster.
\cite{Vitral_2023} reported structural parameters for NGC 6121 through mass-anisotropy Jeans modelling, and found a central mass excess of $\sim$800 M$_{\odot}$, pointing towards the existence of an intermediate-mass black hole or a supercompact black hole population.
Due to its highly eccentric orbit, metal-poor and alpha-enhanced chemistry, very old age, the exotic central content, together with its isolated position in the Galactic similarity context of the manifold, we suggest that NGC 6121 may have an extragalactic origin, possibly being the remnant of an accreted satellite.

\section{Summary and conclusions}
\label{conclusion}

In this paper we have presented a multivariate approach to embed a similarity map of the local volume of the Galaxy, making possible to describe it in terms of orbital, kinematic, chemical and age characteristics, enabling the distinction and isolation of important structures to further study them.
By means of orbit integration of the observed Gaia data, we could validate the Galactic bar as an indispensable feature in the study of the milky way disk, and as an important agent in driving the structure of the spiral arms.
We have shown that the main moving groups arise from orbital resonances and overdensities produced by the bar and spiral arms, as these structures play an important role in star formation.
The radial gradient of metallicity and velocity dispersion could also be mapped with the help of the segragation of populations, showing the heating of an outer disk population as a probable effect of the perturbation by the passage of Sagittarius dSph.
The older populations of the disk, namely the canonical and heated thick disks were clearly defined and isolated as very distinct regions from the thin disk, also showing concentrations in the radial locations of orbital resonances.
Also, the accreted populations that compose the local dynamical halo could be easily mapped due to their highly tilted orbits and chemical features, showing a close connection with the heated thick disk.
The spatial manifold distribution, as a consequence of similarities, enables the visualisation of hierarchical relationships between the disk components and their sub-structures, giving hints about the process of formation of the milky way in a general picture and can even locate star clusters.
We also have proposed that the Na rich population in the inner portion of the Galaxy may be the thin disk consequence of the GSE merger event due to its timescale and chemical patterns.
Moreover, our analysis suggest that the transition zone may have formed from the interactions with merging satellites and their gas infall.

The t-SNE technique has a myriad of applications for the most variate purposes, and its use here can be seen as a proof-of-concept to a whole new possibilities of big data analyses with the upcoming spectroscopic surveys such as 4MOST, as well as the already available LAMOST and RAVE.
Also, for a more detailed analysis, one could isolate the main regions of interest and apply t-SNE individually to each one and obtain even better separations.
The sample can be further expanded to encompass a more complete Galactic volume in exchange for a narrower context of a single chemical survey and a slight loss of radial velocity homogeneity, but still of great value to shed light into the big scenario of galactic evolution.

\section*{Acknowledgements}
EC acknowledges the initial support of CAPES scholarship and the infrastructure of IAG-USP.
This study made extensive use of TOPCAT \citep{topcat} for acquisition, manipulation, cross-match, visualisation and selection of the catalogues. Data was also manipulated besides orbit integration and distance calculation with Python3 \citep{python} and Jupyter notebooks \citep{Kluyver2016jupyter} for testing with t-SNE, and figures elaborated with \textit{Matplotlib} \citep{Hunter_2007}.
We thank Rodrigo Silva, Guilherme Limberg, Amancio Friaça and Angeles Perez-Villegas for the valuable discussions, and the referee that heavily contributed in making this work stronger.

\section*{Data Availability}

The data underlying this article are available in Zenodo, at https://dx.doi.org/10.5281/zenodo.10968535. 
The datasets were derived from sources in the public domain:

- Gaia DR3: https://www.cosmos.esa.int/web/gaia/dr3

- APOGEE DR17: https://www.sdss4.org/dr17/irspec/

- GALAH DR3: https://www.galah-survey.org/dr3/overview/



\bibliographystyle{mnras}
\bibliography{article_main} 

\begin{thebibliography}{}
\makeatletter
\relax
\def\mn@urlcharsother{\let\do\@makeother \do\$\do\&\do\#\do\^\do\_\do\%\do\~}
\def\mn@doi{\begingroup\mn@urlcharsother \@ifnextchar [ {\mn@doi@}
  {\mn@doi@[]}}
\def\mn@doi@[#1]#2{\def\@tempa{#1}\ifx\@tempa\@empty \href
  {http://dx.doi.org/#2} {doi:#2}\else \href {http://dx.doi.org/#2} {#1}\fi
  \endgroup}
\def\mn@eprint#1#2{\mn@eprint@#1:#2::\@nil}
\def\mn@eprint@arXiv#1{\href {http://arxiv.org/abs/#1} {{\tt arXiv:#1}}}
\def\mn@eprint@dblp#1{\href {http://dblp.uni-trier.de/rec/bibtex/#1.xml}
  {dblp:#1}}
\def\mn@eprint@#1:#2:#3:#4\@nil{\def\@tempa {#1}\def\@tempb {#2}\def\@tempc
  {#3}\ifx \@tempc \@empty \let \@tempc \@tempb \let \@tempb \@tempa \fi \ifx
  \@tempb \@empty \def\@tempb {arXiv}\fi \@ifundefined
  {mn@eprint@\@tempb}{\@tempb:\@tempc}{\expandafter \expandafter \csname
  mn@eprint@\@tempb\endcsname \expandafter{\@tempc}}}

\bibitem[\protect\citeauthoryear{{Abdurro'uf} et~al.,}{{Abdurro'uf}
  et~al.}{2022}]{apogeedr17}
{Abdurro'uf} et~al., 2022, \mn@doi [\apjs] {10.3847/1538-4365/ac4414}, \href
  {https://ui.adsabs.harvard.edu/abs/2022ApJS..259...35A} {259, 35}

\bibitem[\protect\citeauthoryear{Aguado et~al.,}{Aguado
  et~al.}{2021}]{Aguado_2021}
Aguado D.~S.,  et~al., 2021, \mn@doi [The Astrophysical Journal Letters]
  {10.3847/2041-8213/abdbb8}, 908, L8

\bibitem[\protect\citeauthoryear{Anders, Chiappini, Santiago, Matijevič,
  Queiroz, Steinmetz  \& Guiglion}{Anders et~al.}{2018}]{anders_2018}
Anders F.,  Chiappini C.,  Santiago B.~X.,  Matijevič G.,  Queiroz A.~B.,
  Steinmetz M.,   Guiglion G.,  2018, \mn@doi [Astronomy \& Astrophysics]
  {10.1051/0004-6361/201833099}, 619, A125

\bibitem[\protect\citeauthoryear{Antoja et~al.,}{Antoja
  et~al.}{2018}]{antoja-helmi}
Antoja T.,  et~al., 2018, \mn@doi [Nature] {10.1038/s41586-018-0510-7}, 561,
  360–362

\bibitem[\protect\citeauthoryear{Asano, Fujii, Baba, Bédorf, Sellentin  \&
  Portegies Zwart}{Asano et~al.}{2020}]{Asano_2020}
Asano T.,  Fujii M.~S.,  Baba J.,  Bédorf J.,  Sellentin E.,
  Portegies Zwart S.,  2020, \mn@doi [Monthly Notices of the Royal
  Astronomical Society] {10.1093/mnras/staa2849}, 499, 2416–2425

\bibitem[\protect\citeauthoryear{Astraatmadja \& Bailer-Jones}{Astraatmadja \&
  Bailer-Jones}{2016}]{abj2018}
Astraatmadja T.~L.,  Bailer-Jones C. A.~L.,  2016, \mn@doi [The Astrophysical
  Journal] {10.3847/0004-637x/832/2/137}, 832, 137

\bibitem[\protect\citeauthoryear{Athanassoula}{Athanassoula}{1980}]{athanassoula1980}
Athanassoula E.,  1980, Astronomy and Astrophysics, 88, 184

\bibitem[\protect\citeauthoryear{{Baade}}{{Baade}}{1944}]{baade1944}
{Baade} W.,  1944, \mn@doi [\apj] {10.1086/144650}, \href
  {https://ui.adsabs.harvard.edu/abs/1944ApJ...100..137B} {100, 137}

\bibitem[\protect\citeauthoryear{{Baade}}{{Baade}}{1951}]{baade1951}
{Baade} W.,  1951, Publications of Michigan Observatory, \href
  {https://ui.adsabs.harvard.edu/abs/1951POMic..10....7B} {10, 7}

\bibitem[\protect\citeauthoryear{Beers, Drilling, Rossi, Chiba, Rhee,
  Führmeister, Norris  \& von Hippel}{Beers et~al.}{2002}]{Beers_2002}
Beers T.~C.,  Drilling J.~S.,  Rossi S.,  Chiba M.,  Rhee J.,  Führmeister B.,
   Norris J.~E.,   von Hippel T.,  2002, \mn@doi [The Astronomical Journal]
  {10.1086/341377}, 124, 931

\bibitem[\protect\citeauthoryear{Belkina, Ciccolella, Anno, Halpert, Spidlen
  \& Snyder-Cappione}{Belkina et~al.}{2019}]{Belkina_2019}
Belkina A.~C.,  Ciccolella C.~O.,  Anno R.,  Halpert R.,  Spidlen J.,
  Snyder-Cappione J.~E.,  2019, \mn@doi [Nature Communications]
  {10.1038/s41467-019-13055-y}, 10, 5415

\bibitem[\protect\citeauthoryear{{Belokurov}, {Erkal}, {Evans}, {Koposov}  \&
  {Deason}}{{Belokurov} et~al.}{2018}]{belokurov2018}
{Belokurov} V.,  {Erkal} D.,  {Evans} N.~W.,  {Koposov} S.~E.,   {Deason}
  A.~J.,  2018, \mn@doi [\mnras] {10.1093/mnras/sty982}, \href
  {https://ui.adsabs.harvard.edu/abs/2018MNRAS.478..611B} {478, 611}

\bibitem[\protect\citeauthoryear{Bernet, Ramos, Antoja, Famaey, Monari, Kazwini
   \& Romero-Gómez}{Bernet et~al.}{2022}]{Bernet2022}
Bernet M.,  Ramos P.,  Antoja T.,  Famaey B.,  Monari G.,  Kazwini H.~A.,
  Romero-Gómez M.,  2022, \mn@doi [Astronomy & Astrophysics]
  {10.1051/0004-6361/202244070}, 667, A116

\bibitem[\protect\citeauthoryear{Boeltzig et~al.,}{Boeltzig
  et~al.}{2016}]{Boeltzig_2016}
Boeltzig A.,  et~al., 2016, \mn@doi [The European Physical Journal A]
  {10.1140/epja/i2016-16075-4}, 52

\bibitem[\protect\citeauthoryear{Bovy}{Bovy}{2015}]{Bovy_2015}
Bovy J.,  2015, \mn@doi [The Astrophysical Journal Supplement Series]
  {10.1088/0067-0049/216/2/29}, 216, 29

\bibitem[\protect\citeauthoryear{Bowen \& Vaughan}{Bowen \&
  Vaughan}{1973}]{Bowen_73}
Bowen I.~S.,  Vaughan A.~H.,  1973, \mn@doi [Appl. Opt.]
  {10.1364/AO.12.001430}, 12, 1430

\bibitem[\protect\citeauthoryear{Buder et~al.,}{Buder et~al.}{2021}]{galah_dr3}
Buder S.,  et~al., 2021, \mn@doi [Monthly Notices of the Royal Astronomical
  Society] {10.1093/mnras/stab1242}, 506, 150

\bibitem[\protect\citeauthoryear{Carrillo, Hawkins, Jofré, de Brito Silva,
  Das  \& Lucey}{Carrillo et~al.}{2022}]{Carrillo_2022}
Carrillo A.,  Hawkins K.,  Jofré P.,  de Brito Silva D.,  Das P.,   Lucey
  M.,  2022, \mn@doi [Monthly Notices of the Royal Astronomical Society]
  {10.1093/mnras/stac518}, 513, 1557–1580

\bibitem[\protect\citeauthoryear{Chen, Zhao  \& Zhang}{Chen
  et~al.}{2022}]{Chen_2022}
Chen Y.,  Zhao G.,   Zhang H.,  2022, \mn@doi [The Astrophysical Journal
  Letters] {10.3847/2041-8213/ac898e}, 936, L7

\bibitem[\protect\citeauthoryear{Chiappini, Matteucci  \& Gratton}{Chiappini
  et~al.}{1997}]{chiappini1997}
Chiappini C.,  Matteucci F.,   Gratton R.,  1997, \mn@doi [The Astrophysical
  Journal] {10.1086/303726}, 477, 765–780

\bibitem[\protect\citeauthoryear{Dai \& Tong}{Dai \&
  Tong}{2018}]{dai2018visualizing}
Dai J.-M.,  Tong J.,  2018, Visualizing the Hidden Features of Galaxy
  Morphology with Machine Learning (\mn@eprint {arXiv} {1807.05657})

\bibitem[\protect\citeauthoryear{Das, Hawkins  \& Jofré}{Das
  et~al.}{2020}]{Das2020}
Das P.,  Hawkins K.,   Jofré P.,  2020, \mn@doi [Monthly Notices of the Royal
  Astronomical Society] {10.1093/mnras/stz3537}, 493, 5195–5207

\bibitem[\protect\citeauthoryear{{Das}, {Huang}, {Ciuc{\u{a}}}  \&
  {Fragkoudi}}{{Das} et~al.}{2024}]{das2024}
{Das} P.,  {Huang} Y.,  {Ciuc{\u{a}}} I.,   {Fragkoudi} F.,  2024, \mn@doi
  [\mnras] {10.1093/mnras/stad3344}, \href
  {https://ui.adsabs.harvard.edu/abs/2024MNRAS.527.4505D} {527, 4505}

\bibitem[\protect\citeauthoryear{{De Silva} et~al.,}{{De Silva}
  et~al.}{2015}]{desilva2015}
{De Silva} G.~M.,  et~al., 2015, \mn@doi [\mnras] {10.1093/mnras/stv327}, \href
  {https://ui.adsabs.harvard.edu/abs/2015MNRAS.449.2604D} {449, 2604}

\bibitem[\protect\citeauthoryear{Deason, Belokurov, Evans  \& Johnston}{Deason
  et~al.}{2013}]{deason2013}
Deason A.~J.,  Belokurov V.,  Evans N.~W.,   Johnston K.~V.,  2013, \mn@doi
  [The Astrophysical Journal] {10.1088/0004-637x/763/2/113}, 763, 113

\bibitem[\protect\citeauthoryear{Di~Matteo, Haywood, Lehnert, Katz, Khoperskov,
  Snaith, Gómez  \& Robichon}{Di~Matteo et~al.}{2019}]{DiMatteo_2019}
Di~Matteo P.,  Haywood M.,  Lehnert M.~D.,  Katz D.,  Khoperskov S.,  Snaith
  O.~N.,  Gómez A.,   Robichon N.,  2019, \mn@doi [Astronomy & Astrophysics]
  {10.1051/0004-6361/201834929}, 632, A4

\bibitem[\protect\citeauthoryear{Dias, Monteiro, Lépine  \& Barros}{Dias
  et~al.}{2019}]{dias2019}
Dias W.~S.,  Monteiro H.,  Lépine J. R.~D.,   Barros D.~A.,  2019, \mn@doi
  [Monthly Notices of the Royal Astronomical Society] {10.1093/mnras/stz1196},
  486, 5726–5736

\bibitem[\protect\citeauthoryear{{Erdem}}{{Erdem}}{2020}]{tds_tsne}
{Erdem} K.,  2020, t-SNE clearly explained,
  \url{https://towardsdatascience.com/t-sne-clearly-explained-d84c537f53a}

\bibitem[\protect\citeauthoryear{Feuillet, Sahlholdt, Feltzing  \&
  Casagrande}{Feuillet et~al.}{2021}]{Feuillet_2021}
Feuillet D.~K.,  Sahlholdt C.~L.,  Feltzing S.,   Casagrande L.,  2021, \mn@doi
  [Monthly Notices of the Royal Astronomical Society] {10.1093/mnras/stab2614},
  508, 1489

\bibitem[\protect\citeauthoryear{Feuillet, Feltzing, Sahlholdt  \&
  Bensby}{Feuillet et~al.}{2022}]{Feuillet_2022}
Feuillet D.~K.,  Feltzing S.,  Sahlholdt C.,   Bensby T.,  2022, \mn@doi [The
  Astrophysical Journal] {10.3847/1538-4357/ac76ba}, 934, 21

\bibitem[\protect\citeauthoryear{Forbes}{Forbes}{2020}]{Forbes_2020}
Forbes D.~A.,  2020, \mn@doi [Monthly Notices of the Royal Astronomical
  Society] {10.1093/mnras/staa245}, 493, 847–854

\bibitem[\protect\citeauthoryear{{Gaia Collaboration} et~al.,}{{Gaia
  Collaboration} et~al.}{2016}]{gaia_main}
{Gaia Collaboration} et~al., 2016, \mn@doi [A&A] {10.1051/0004-6361/201629272},
  595, A1

\bibitem[\protect\citeauthoryear{{Gaia Collaboration} et~al.,}{{Gaia
  Collaboration} et~al.}{2018a}]{gaiaHR_2018}
{Gaia Collaboration} et~al., 2018a, \mn@doi [Astronomy \& Astrophysics]
  {10.1051/0004-6361/201832843}, 616, A10

\bibitem[\protect\citeauthoryear{{Gaia Collaboration} et~al.,}{{Gaia
  Collaboration} et~al.}{2018b}]{katz2018}
{Gaia Collaboration} et~al., 2018b, \mn@doi [\aap]
  {10.1051/0004-6361/201832865}, \href
  {https://ui.adsabs.harvard.edu/abs/2018A&A...616A..11G} {616, A11}

\bibitem[\protect\citeauthoryear{{Gaia Collaboration} et~al.,}{{Gaia
  Collaboration} et~al.}{2023a}]{gaia_dr3}
{Gaia Collaboration} et~al., 2023a, \mn@doi [\aap]
  {10.1051/0004-6361/202243940}, \href
  {https://ui.adsabs.harvard.edu/abs/2023A&A...674A...1G} {674, A1}

\bibitem[\protect\citeauthoryear{{Gaia Collaboration} et~al.,}{{Gaia
  Collaboration} et~al.}{2023b}]{drimmel_2023}
{Gaia Collaboration} et~al., 2023b, \mn@doi [\aap]
  {10.1051/0004-6361/202243797}, \href
  {https://ui.adsabs.harvard.edu/abs/2023A&A...674A..37G} {674, A37}

\bibitem[\protect\citeauthoryear{García~Pérez et~al.,}{García~Pérez
  et~al.}{2016}]{GarciaPerez_2016}
García~Pérez A.~E.,  et~al., 2016, \mn@doi [The Astronomical Journal]
  {10.3847/0004-6256/151/6/144}, 151, 144

\bibitem[\protect\citeauthoryear{Garma-Oehmichen, Martinez-Medina,
  Hernández-Toledo  \& Puerari}{Garma-Oehmichen et~al.}{2021}]{garma2021}
Garma-Oehmichen L.,  Martinez-Medina L.,  Hernández-Toledo H.,   Puerari I.,
  2021, \mn@doi [Monthly Notices of the Royal Astronomical Society]
  {10.1093/mnras/stab333}, 502, 4708–4722

\bibitem[\protect\citeauthoryear{{Gilmore} \& {Reid}}{{Gilmore} \&
  {Reid}}{1983}]{gilmorereid1983}
{Gilmore} G.,  {Reid} N.,  1983, \mn@doi [\mnras] {10.1093/mnras/202.4.1025},
  \href {https://ui.adsabs.harvard.edu/abs/1983MNRAS.202.1025G} {202, 1025}

\bibitem[\protect\citeauthoryear{Gilmore, Wyse  \& Kuijken}{Gilmore
  et~al.}{1989}]{Gilmore_1989}
Gilmore G.,  Wyse R. F.~G.,   Kuijken K.,  1989, \mn@doi [Annual Review of
  Astronomy and Astrophysics] {10.1146/annurev.aa.27.090189.003011}, 27,
  555–627

\bibitem[\protect\citeauthoryear{Giribaldi \& Smiljanic}{Giribaldi \&
  Smiljanic}{2023}]{Giribaldi_2023}
Giribaldi R.~E.,  Smiljanic R.,  2023, \mn@doi [Astronomy &amp; Astrophysics]
  {10.1051/0004-6361/202245404}, 673, A18

\bibitem[\protect\citeauthoryear{{Gunn} et~al.,}{{Gunn} et~al.}{2006a}]{sloan}
{Gunn} J.~E.,  et~al., 2006a, \mn@doi [\aj] {10.1086/500975}, \href
  {https://ui.adsabs.harvard.edu/abs/2006AJ....131.2332G} {131, 2332}

\bibitem[\protect\citeauthoryear{Gunn et~al.,}{Gunn et~al.}{2006b}]{Gunn_2006}
Gunn J.~E.,  et~al., 2006b, \mn@doi [The Astronomical Journal]
  {10.1086/500975}, 131, 2332–2359

\bibitem[\protect\citeauthoryear{Hawkins, Jofr{\'{e} }, Masseron  \&
  Gilmore}{Hawkins et~al.}{2015}]{Hawkins_2015}
Hawkins K.,  Jofr{\'{e} } P.,  Masseron T.,   Gilmore G.,  2015, \mn@doi
  [Monthly Notices of the Royal Astronomical Society] {10.1093/mnras/stv1586},
  453, 758

\bibitem[\protect\citeauthoryear{Hayes et~al.,}{Hayes
  et~al.}{2018}]{Hayes_2018}
Hayes C.~R.,  et~al., 2018, \mn@doi [The Astrophysical Journal]
  {10.3847/1538-4357/aa9cec}, 852, 49

\bibitem[\protect\citeauthoryear{Haywood, Matteo, Lehnert, Snaith, Khoperskov
  \& G{\'{o} }mez}{Haywood et~al.}{2018a}]{haywood2018}
Haywood M.,  Matteo P.~D.,  Lehnert M.~D.,  Snaith O.,  Khoperskov S.,
  G{\'{o} }mez A.,  2018a, \mn@doi [The Astrophysical Journal]
  {10.3847/1538-4357/aad235}, 863, 113

\bibitem[\protect\citeauthoryear{Haywood, Matteo, Lehnert, Snaith, Khoperskov
  \& G{\'{o} }mez}{Haywood et~al.}{2018b}]{Haywood_2018}
Haywood M.,  Matteo P.~D.,  Lehnert M.~D.,  Snaith O.,  Khoperskov S.,
  G{\'{o} }mez A.,  2018b, \mn@doi [The Astrophysical Journal]
  {10.3847/1538-4357/aad235}, 863, 113

\bibitem[\protect\citeauthoryear{{Helmi}, {White}, {de Zeeuw}  \&
  {Zhao}}{{Helmi} et~al.}{1999}]{helmi1999}
{Helmi} A.,  {White} S. D.~M.,  {de Zeeuw} P.~T.,   {Zhao} H.,  1999, \mn@doi
  [\nat] {10.1038/46980}, \href
  {https://ui.adsabs.harvard.edu/abs/1999Natur.402...53H} {402, 53}

\bibitem[\protect\citeauthoryear{Helmi, Babusiaux, Koppelman, Massari,
  Veljanoski  \& Brown}{Helmi et~al.}{2018}]{helmi2018}
Helmi A.,  Babusiaux C.,  Koppelman H.~H.,  Massari D.,  Veljanoski J.,   Brown
  A. G.~A.,  2018, \mn@doi [Nature] {10.1038/s41586-018-0625-x}, 563, 85–88

\bibitem[\protect\citeauthoryear{House et~al.,}{House
  et~al.}{2023}]{House_2023}
House L.~R.,  et~al., 2023, \mn@doi [The Astrophysical Journal]
  {10.3847/1538-4357/accdd0}, 950, 82

\bibitem[\protect\citeauthoryear{{Hubble}}{{Hubble}}{1925}]{hubble1925}
{Hubble} E.~P.,  1925, The Observatory, \href
  {https://ui.adsabs.harvard.edu/abs/1925Obs....48..139H} {48, 139}

\bibitem[\protect\citeauthoryear{Hunter}{Hunter}{2007}]{Hunter_2007}
Hunter J.~D.,  2007, \mn@doi [Computing in Science \& Engineering]
  {10.1109/MCSE.2007.55}, 9, 90

\bibitem[\protect\citeauthoryear{{Ibata}, {Gilmore}  \& {Irwin}}{{Ibata}
  et~al.}{1994}]{ibata1994}
{Ibata} R.~A.,  {Gilmore} G.,   {Irwin} M.~J.,  1994, \mn@doi [\nat]
  {10.1038/370194a0}, \href
  {https://ui.adsabs.harvard.edu/abs/1994Natur.370..194I} {370, 194}

\bibitem[\protect\citeauthoryear{Jönsson et~al.,}{Jönsson
  et~al.}{2020}]{Jonsson_2020}
Jönsson H.,  et~al., 2020, \mn@doi [The Astronomical Journal]
  {10.3847/1538-3881/aba592}, 160, 120

\bibitem[\protect\citeauthoryear{Karakas, Lugaro, Carlos, Cseh, Kamath  \&
  García-Hernández}{Karakas et~al.}{2018}]{karakas2018}
Karakas A.~I.,  Lugaro M.,  Carlos M.,  Cseh B.,  Kamath D.,
  García-Hernández D.~A.,  2018, \mn@doi [Monthly Notices of the Royal
  Astronomical Society] {10.1093/mnras/sty625}, 477, 421–437

\bibitem[\protect\citeauthoryear{Katz et~al.,}{Katz et~al.}{2023}]{Katz_2023}
Katz D.,  et~al., 2023, \mn@doi [Astronomy and Astrophysics]
  {10.1051/0004-6361/202244220}, 674, A5

\bibitem[\protect\citeauthoryear{Khoperskov \& Gerhard}{Khoperskov \&
  Gerhard}{2022}]{Khoperskov_2022}
Khoperskov S.,  Gerhard O.,  2022, \mn@doi [Astronomy and Astrophysics]
  {10.1051/0004-6361/202141836}, 663, A38

\bibitem[\protect\citeauthoryear{Kluyver et~al.,}{Kluyver
  et~al.}{2016}]{Kluyver2016jupyter}
Kluyver T.,  et~al., 2016, in Loizides F.,  Schmidt B.,  eds, Positioning and
  Power in Academic Publishing: Players, Agents and Agendas. pp 87 -- 90

\bibitem[\protect\citeauthoryear{Kobayashi \& Nomoto}{Kobayashi \&
  Nomoto}{2009}]{Kobayashi_2009}
Kobayashi C.,  Nomoto K.,  2009, \mn@doi [The Astrophysical Journal]
  {10.1088/0004-637x/707/2/1466}, 707, 1466

\bibitem[\protect\citeauthoryear{Kobayashi, Karakas  \& Lugaro}{Kobayashi
  et~al.}{2020}]{Kobayashi_2020}
Kobayashi C.,  Karakas A.~I.,   Lugaro M.,  2020, \mn@doi [The Astrophysical
  Journal] {10.3847/1538-4357/abae65}, 900, 179

\bibitem[\protect\citeauthoryear{Leung \& Bovy}{Leung \&
  Bovy}{2018}]{Leung_2018}
Leung H.~W.,  Bovy J.,  2018, \mn@doi [Monthly Notices of the Royal
  Astronomical Society] {10.1093/mnras/sty3217}

\bibitem[\protect\citeauthoryear{{Lindblad}}{{Lindblad}}{1927}]{lindblad1926}
{Lindblad} B.,  1927, Meddelanden fran Astronomiska Observatorium Uppsala, 13

\bibitem[\protect\citeauthoryear{Lucey, Pearson, Hunt, Hawkins, Ness, Petersen,
  Price-Whelan  \& Weinberg}{Lucey et~al.}{2023}]{Lucey_2023}
Lucey M.,  Pearson S.,  Hunt J. A.~S.,  Hawkins K.,  Ness M.,  Petersen M.~S.,
  Price-Whelan A.~M.,   Weinberg M.~D.,  2023, \mn@doi [Monthly Notices of the
  Royal Astronomical Society] {10.1093/mnras/stad406}, 520, 4779

\bibitem[\protect\citeauthoryear{Mackereth et~al.,}{Mackereth
  et~al.}{2019}]{Mackereth_2019}
Mackereth J.~T.,  et~al., 2019, \mn@doi [Monthly Notices of the Royal
  Astronomical Society] {10.1093/mnras/stz1521}, 489, 176

\bibitem[\protect\citeauthoryear{Magrini et~al.,}{Magrini
  et~al.}{2018}]{Magrini_2018}
Magrini L.,  et~al., 2018, \mn@doi [Astronomy and Astrophysics]
  {10.1051/0004-6361/201832841}, 617, A106

\bibitem[\protect\citeauthoryear{Majewski et~al.,}{Majewski
  et~al.}{2017}]{apogee}
Majewski S.~R.,  et~al., 2017, \mn@doi [The Astronomical Journal]
  {10.3847/1538-3881/aa784d}, 154, 94

\bibitem[\protect\citeauthoryear{Marigo et~al.,}{Marigo
  et~al.}{2017}]{Marigo_2017}
Marigo P.,  et~al., 2017, \mn@doi [The Astrophysical Journal]
  {10.3847/1538-4357/835/1/77}, 835, 77

\bibitem[\protect\citeauthoryear{Montalb{\'{a}}n et~al.,}{Montalb{\'{a}}n
  et~al.}{2021}]{Montalb_n_2021}
Montalb{\'{a}}n J.,  et~al., 2021, \mn@doi [Nature Astronomy]
  {10.1038/s41550-021-01347-7}, 5, 640

\bibitem[\protect\citeauthoryear{Myeong, Vasiliev, Iorio, Evans  \&
  Belokurov}{Myeong et~al.}{2019}]{Myeong_2019}
Myeong G.~C.,  Vasiliev E.,  Iorio G.,  Evans N.~W.,   Belokurov V.,  2019,
  \mn@doi [Monthly Notices of the Royal Astronomical Society]
  {10.1093/mnras/stz1770}, 488, 1235

\bibitem[\protect\citeauthoryear{Myeong, Belokurov, Aguado, Evans, Caldwell  \&
  Bradley}{Myeong et~al.}{2022}]{Myeong_2022}
Myeong G.~C.,  Belokurov V.,  Aguado D.~S.,  Evans N.~W.,  Caldwell N.,
  Bradley J.,  2022, \mn@doi [The Astrophysical Journal]
  {10.3847/1538-4357/ac8d68}, 938, 21

\bibitem[\protect\citeauthoryear{Naidu, Conroy, Bonaca, Johnson, Ting,
  Caldwell, Zaritsky  \& Cargile}{Naidu et~al.}{2020}]{Naidu_2020}
Naidu R.~P.,  Conroy C.,  Bonaca A.,  Johnson B.~D.,  Ting Y.-S.,  Caldwell N.,
   Zaritsky D.,   Cargile P.~A.,  2020, \mn@doi [The Astrophysical Journal]
  {10.3847/1538-4357/abaef4}, 901, 48

\bibitem[\protect\citeauthoryear{Nandakumar et~al.,}{Nandakumar
  et~al.}{2022}]{Nandakumar_2022}
Nandakumar G.,  et~al., 2022, \mn@doi [Monthly Notices of the Royal
  Astronomical Society] {10.1093/mnras/stac873}, 513, 232

\bibitem[\protect\citeauthoryear{{Nissen} \& {Schuster}}{{Nissen} \&
  {Schuster}}{2010}]{nissen2010}
{Nissen} P.~E.,  {Schuster} W.~J.,  2010, \mn@doi [\aap]
  {10.1051/0004-6361/200913877}, \href
  {https://ui.adsabs.harvard.edu/abs/2010A&A...511L..10N} {511, L10}

\bibitem[\protect\citeauthoryear{{Nordlander}, {Gruyters}, {Richard}  \&
  {Korn}}{{Nordlander} et~al.}{2024}]{nordlander_2024}
{Nordlander} T.,  {Gruyters} P.,  {Richard} O.,   {Korn} A.~J.,  2024, \mn@doi
  [\mnras] {10.1093/mnras/stad3973}, \href
  {https://ui.adsabs.harvard.edu/abs/2024MNRAS.527.12120} {527, 12120}

\bibitem[\protect\citeauthoryear{{Oort}}{{Oort}}{1927}]{oort1927}
{Oort} J.~H.,  1927, \bain, \href
  {https://ui.adsabs.harvard.edu/abs/1927BAN.....3..275O} {3, 275}

\bibitem[\protect\citeauthoryear{Pettitt, Ragan  \& Smith}{Pettitt
  et~al.}{2019}]{pettitt2019}
Pettitt A.~R.,  Ragan S.~E.,   Smith M.~C.,  2019, \mn@doi [Monthly Notices of
  the Royal Astronomical Society] {10.1093/mnras/stz3155}

\bibitem[\protect\citeauthoryear{Pinsonneault et~al.,}{Pinsonneault
  et~al.}{2018}]{Pinsonneault_2018}
Pinsonneault M.~H.,  et~al., 2018, \mn@doi [The Astrophysical Journal
  Supplement Series] {10.3847/1538-4365/aaebfd}, 239, 32

\bibitem[\protect\citeauthoryear{Piskunov \& Valenti}{Piskunov \&
  Valenti}{2016}]{Piskunov_2016}
Piskunov N.,  Valenti J.~A.,  2016, \mn@doi [Astronomy &amp; Astrophysics]
  {10.1051/0004-6361/201629124}, 597, A16

\bibitem[\protect\citeauthoryear{Poli{\v c}ar, Stra{\v z}ar  \& Zupan}{Poli{\v
  c}ar et~al.}{2019}]{open_tsne}
Poli{\v c}ar P.~G.,  Stra{\v z}ar M.,   Zupan B.,  2019, \mn@doi [bioRxiv]
  {10.1101/731877}

\bibitem[\protect\citeauthoryear{Pérez-Villegas, Portail, Wegg  \&
  Gerhard}{Pérez-Villegas et~al.}{2017}]{PerezVillegas_2017}
Pérez-Villegas A.,  Portail M.,  Wegg C.,   Gerhard O.,  2017, \mn@doi [The
  Astrophysical Journal] {10.3847/2041-8213/aa6c26}, 840, L2

\bibitem[\protect\citeauthoryear{Queiroz et~al.,}{Queiroz
  et~al.}{2023}]{Queiroz_2023}
Queiroz A. B.~A.,  et~al., 2023, \mn@doi [Astronomy and Astrophysics]
  {10.1051/0004-6361/202245399}, 673, A155

\bibitem[\protect\citeauthoryear{Recio-Blanco et~al.,}{Recio-Blanco
  et~al.}{2014}]{Recio_Blanco_2014}
Recio-Blanco A.,  et~al., 2014, \mn@doi [Astronomy and Astrophysics]
  {10.1051/0004-6361/201322944}, 567, A5

\bibitem[\protect\citeauthoryear{{Schwarzschild}}{{Schwarzschild}}{1952}]{schwarzschild1952}
{Schwarzschild} M.,  1952, \mn@doi [\aj] {10.1086/106710}, \href
  {https://ui.adsabs.harvard.edu/abs/1952AJ.....57...57S} {57, 57}

\bibitem[\protect\citeauthoryear{Schönrich}{Schönrich}{2012}]{schonrich2012}
Schönrich R.,  2012, \mn@doi [Monthly Notices of the Royal Astronomical
  Society] {10.1111/j.1365-2966.2012.21631.x}, 427, 274–287

\bibitem[\protect\citeauthoryear{Sharma et~al.,}{Sharma
  et~al.}{2017}]{Sharma_2017}
Sharma S.,  et~al., 2017, \mn@doi [Monthly Notices of the Royal Astronomical
  Society] {10.1093/mnras/stx2582}, 473, 2004

\bibitem[\protect\citeauthoryear{{Sheinis} et~al.,}{{Sheinis}
  et~al.}{2015}]{Sheinis_2015}
{Sheinis} A.,  et~al., 2015, \mn@doi [Journal of Astronomical Telescopes,
  Instruments, and Systems] {10.1117/1.JATIS.1.3.035002}, \href
  {https://ui.adsabs.harvard.edu/abs/2015JATIS...1c5002S} {1, 035002}

\bibitem[\protect\citeauthoryear{Smiljanic et~al.,}{Smiljanic
  et~al.}{2016}]{smiljanic2016}
Smiljanic R.,  et~al., 2016, \mn@doi [Astronomy and Astrophysics]
  {10.1051/0004-6361/201528014}, 589, A115

\bibitem[\protect\citeauthoryear{Smith et~al.,}{Smith
  et~al.}{2021}]{Smith_2021}
Smith V.~V.,  et~al., 2021, \mn@doi [The Astronomical Journal]
  {10.3847/1538-3881/abefdc}, 161, 254

\bibitem[\protect\citeauthoryear{{Song}, {Li}, {Wu}, {Pietrinferni}, {Poon}  \&
  {Xie}}{{Song} et~al.}{2018}]{song_2018}
{Song} F.,  {Li} Y.,  {Wu} T.,  {Pietrinferni} A.,  {Poon} H.,   {Xie} Y.,
  2018, \mn@doi [\apj] {10.3847/1538-4357/aaecd3}, \href
  {https://ui.adsabs.harvard.edu/abs/2018ApJ...869..109S} {869, 109}

\bibitem[\protect\citeauthoryear{Spina et~al.,}{Spina
  et~al.}{2017}]{Spina_2017}
Spina L.,  et~al., 2017, \mn@doi [Monthly Notices of the Royal Astronomical
  Society] {10.1093/mnras/stx2938}

\bibitem[\protect\citeauthoryear{Springel, Frenk  \& White}{Springel
  et~al.}{2006}]{Springel_2006}
Springel V.,  Frenk C.~S.,   White S. D.~M.,  2006, \mn@doi [Nature]
  {10.1038/nature04805}, 440, 1137–1144

\bibitem[\protect\citeauthoryear{Steinhardt, Weaver, Maxfield, Davidzon,
  Faisst, Masters, Schemel  \& Toft}{Steinhardt et~al.}{2020}]{Steinhardt_2020}
Steinhardt C.~L.,  Weaver J.~R.,  Maxfield J.,  Davidzon I.,  Faisst A.~L.,
  Masters D.,  Schemel M.,   Toft S.,  2020, \mn@doi [The Astrophysical
  Journal] {10.3847/1538-4357/ab76be}, 891, 136

\bibitem[\protect\citeauthoryear{Sun, Shen, Jiang  \& Liu}{Sun
  et~al.}{2024}]{sun2024mapping}
Sun W.,  Shen H.,  Jiang B.,   Liu X.,  2024, Mapping the Chemo-dynamics of the
  Galactic disk using the LAMOST and APOGEE red clump stars (\mn@eprint {arXiv}
  {2403.01842})

\bibitem[\protect\citeauthoryear{{Taylor}}{{Taylor}}{2005}]{topcat}
{Taylor} M.~B.,  2005, in {Shopbell} P.,  {Britton} M.,   {Ebert} R.,  eds,
  Astronomical Society of the Pacific Conference Series Vol. 347, Astronomical
  Data Analysis Software and Systems XIV. p.~29

\bibitem[\protect\citeauthoryear{Traven et~al.,}{Traven
  et~al.}{2017}]{Traven_2017}
Traven G.,  et~al., 2017, \mn@doi [The Astrophysical Journal Supplement Series]
  {10.3847/1538-4365/228/2/24}, 228, 24

\bibitem[\protect\citeauthoryear{Van~Rossum \& Drake}{Van~Rossum \&
  Drake}{2009}]{python}
Van~Rossum G.,  Drake F.~L.,  2009, Python 3 Reference Manual.
CreateSpace, Scotts Valley, CA

\bibitem[\protect\citeauthoryear{Vitral, Libralato, Kremer, Mamon, Bellini,
  Bedin  \& Anderson}{Vitral et~al.}{2023}]{Vitral_2023}
Vitral E.,  Libralato M.,  Kremer K.,  Mamon G.~A.,  Bellini A.,  Bedin L.~R.,
   Anderson J.,  2023, \mn@doi [Monthly Notices of the Royal Astronomical
  Society] {10.1093/mnras/stad1068}, 522, 5740–5757

\bibitem[\protect\citeauthoryear{{Vyssotsky}}{{Vyssotsky}}{1951}]{vyssotsky1951}
{Vyssotsky} A.~N.,  1951, \mn@doi [\aj] {10.1086/106515}, \href
  {https://ui.adsabs.harvard.edu/abs/1951AJ.....56...62V} {56, 62}

\bibitem[\protect\citeauthoryear{Wattenberg, Viégas  \& Johnson}{Wattenberg
  et~al.}{2016}]{wattenberg2016how}
Wattenberg M.,  Viégas F.,   Johnson I.,  2016, \mn@doi [Distill]
  {10.23915/distill.00002}

\bibitem[\protect\citeauthoryear{{Wilson} et~al.,}{{Wilson}
  et~al.}{2019a}]{apogeespec}
{Wilson} J.~C.,  et~al., 2019a, \mn@doi [\pasp] {10.1088/1538-3873/ab0075},
  \href {https://ui.adsabs.harvard.edu/abs/2019PASP..131e5001W} {131, 055001}

\bibitem[\protect\citeauthoryear{{Wilson} et~al.,}{{Wilson}
  et~al.}{2019b}]{Wilson_2019}
{Wilson} J.~C.,  et~al., 2019b, \mn@doi [\pasp] {10.1088/1538-3873/ab0075},
  \href {https://ui.adsabs.harvard.edu/abs/2019PASP..131e5001W} {131, 055001}

\bibitem[\protect\citeauthoryear{Zinn et~al.,}{Zinn et~al.}{2022}]{Zinn_2022}
Zinn J.~C.,  et~al., 2022, \mn@doi [The Astrophysical Journal]
  {10.3847/1538-4357/ac2c83}, 926, 191

\bibitem[\protect\citeauthoryear{da Silva \& Smiljanic}{da~Silva \&
  Smiljanic}{2023}]{da_Silva_2023}
da Silva A.~R.,  Smiljanic R.,  2023, \mn@doi [Astronomy &amp; Astrophysics]
  {10.1051/0004-6361/202347229}, 677, A74

\bibitem[\protect\citeauthoryear{{de Souza} \& {Teixeira}}{{de Souza} \&
  {Teixeira}}{2007}]{rama_2007}
{de Souza} R.~E.,  {Teixeira} R.,  2007, \mn@doi [\aap]
  {10.1051/0004-6361:20066257}, \href
  {https://ui.adsabs.harvard.edu/abs/2007A&A...471..475D} {471, 475}

\bibitem[\protect\citeauthoryear{{deVaucouleurs}}{{deVaucouleurs}}{1964}]{deVaucouleurs1964}
{deVaucouleurs} G.,  1964, in The Galaxy and the Magellanic Clouds. p.~195

\bibitem[\protect\citeauthoryear{van~der Maaten \& Hinton}{van~der Maaten \&
  Hinton}{2008}]{tsne}
van~der Maaten L.,  Hinton G.,  2008, Journal of Machine Learning Research, 9,
  2579

\makeatother
\end{thebibliography}




\appendix

\section{Selection polygons of CM and Kiel diagrams}
\label{poly_append}

The polygons used for the RGB and MS subsample selection are shown in this appendix as a list of vertex coordinate pairs for the RGB cuts in Table \ref{giants_polygons}, and for the MS cuts in Table \ref{mainseq_polygons}.

\begin{table}
\caption{Vertex list of the polygons drawn for the selection cuts of the red giant branch in the CM and Kiel diagrams for both APOGEE and GALAH. BP-RP and G are the Gaia magnitudes, and units of Teff and log G are in K and dex, respectively.}
\begin{center}
\begin{tabular}{cc}
\hline\hline
CM cut & Kiel cut \\
(BP-RP, G) & (Teff, log G) \\
\cmidrule(lr){1-1} \cmidrule(lr){2-2}
(0.92, 3.90)  & (3449, 0.01) \\
(1.11, 4.30)  & (3431, 0.84) \\
(1.85, 5.20)  & (3758, 1.52) \\
(3.08, 5.75)  & (3839, 1.63) \\
(4.12, 5.39)  & (4311, 2.57) \\
(3.66, -3.04) & (4511, 2.91) \\
(1.39, -4.37) & (4792, 3.64) \\
(0.97, -3.59) & (5037, 4.18) \\
(0.81, 0.00)  & (5436, 4.21) \\
(0.88, 2.90)  & (5653, 4.18) \\
              & (5726, 3.89) \\
              & (5681, 3.48) \\
              & (5490, 2.88) \\
              & (5472, 2.36) \\
              & (5508, 2.13) \\
              & (5381, 1.96) \\
              & (5109, 1.66) \\
              & (4674, 0.76) \\
              & (3885, 0.01) \\
\hline\hline
\end{tabular}
\end{center}
\label{giants_polygons}
\end{table}

\begin{table}
\caption{Vertex list of the polygons drawn for the selection cuts of the main sequence in the CM and Kiel diagrams for APOGEE and GALAH separately. Only the second and third points have differences in the Kiel polygon. BP-RP and G are the Gaia magnitudes, and units of Teff and log G are in K and dex, respectively.}
\begin{center}
\begin{tabular}{ccc}
\hline\hline
CM cut & Kiel cut APOGEE & Kiel cut GALAH \\
(BP-RP, G) & (Teff, log G) & (Teff, log G) \\
\cmidrule(lr){1-1} \cmidrule(lr){2-2} \cmidrule(lr){3-3}
(0.62, -0.15)  &  (3537, 4.45)  &  (3537, 4.45)  \\
(0.73, 1.08)   &  (5700, 4.36)  &  (5500, 4.23)  \\
(0.88, 3.28)   &  (6300, 4.20)  &  (6280, 4.10)  \\
(0.89, 3.85)   &  (6360, 4.00)  &  (6360, 4.00)  \\
(1.05, 4.88)   &  (5762, 3.35)  &  (5762, 3.35)  \\
(1.64, 6.36)   &  (5989, 3.20)  &  (5989, 3.20)  \\
(3.11, 9.67)   &  (6443, 3.30)  &  (6443, 3.30)  \\
(2.22, 10.75)  &  (6972, 3.30)  &  (6972, 3.30)  \\
(1.48, 10.21)  &  (8131, 3.50)  &  (8131, 3.50)  \\
(0.67, 6.96)   &  (8131, 4.68)  &  (8131, 4.68)  \\
(0.12, 3.55)   &  (6972, 4.68)  &  (6972, 4.68)  \\
(-0.07, -0.57) &  (6421, 4.83)  &  (6421, 4.83)  \\
(0.46, -1.18)  &  (5924, 4.68)  &  (5924, 4.68)  \\
               &  (3494, 4.85)  &  (3494, 4.85)  \\
\hline\hline
\end{tabular}
\end{center}
\label{mainseq_polygons}
\end{table}

\section{Age analysis}
\label{ages}

Here we show comparisons between age estimations of APOGEE and GALAH together with other related catalogues of ages with the same determination methods.
Also, we present comparisons and some insights about the use of chemical clocks in age estimation, their relation with astroNN and BSTEP, and their manifold behaviour compared to BSTEP ages.

\begin{figure}
	\includegraphics[width=0.49\columnwidth]{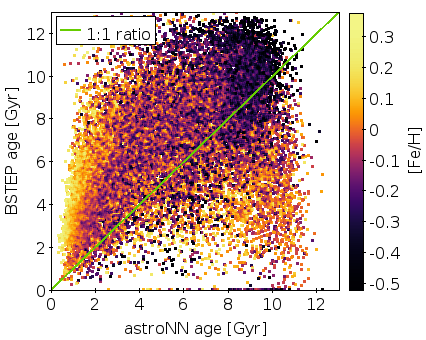}
	\includegraphics[width=0.49\columnwidth]{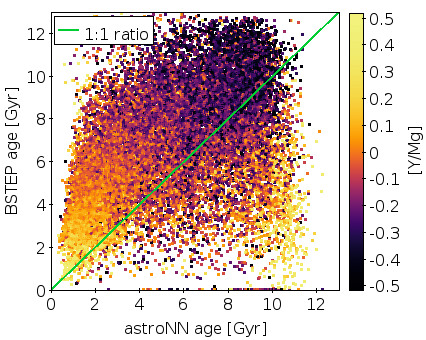}
    \caption{BSTEP ages vs astroNN ages, colour coded by [Fe/H] in the left panel and by [Y/Mg] \citep{Spina_2017} in the right panel.}
    \label{astronn_bstep}
\end{figure}

\cite{Spina_2017} found a chemical clock based on [Y/Mg] ratios to determine solar-twin ages, which can be used together with [Fe/H] as a guide to inspect the behaviour of the BSTEP x astroNN age relation, and more generally bayesian-isochrone vs asteroseismology estimators.
Figure \ref{astronn_bstep} shows the ages of common stars between APOGEE and GALAH.

Although there is some correlation between the ages ($r_{Pearson}=0.4$), there is a knee-like trend with scattered metal-rich stars at older ages, probably originating from the different methods (isochrone-based and asteroseismology) and two different samples (\citealt{Montalb_n_2021} for low-metallicity stars and APOKASC-2 from \citealt{Pinsonneault_2018}) in the APOGEE estimation.
This large deviation of metal-rich stars from the 1:1 ratio of astroNN x BSTEP, signals that one must be cautious when using the ages at this metallicities.

\begin{figure}
	\includegraphics[width=0.46\columnwidth]{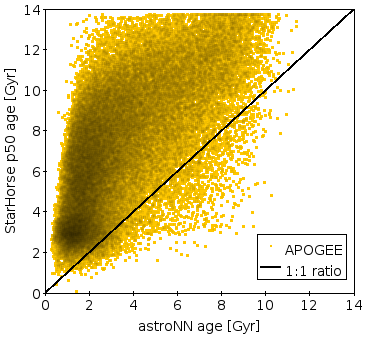}
	\includegraphics[width=0.52\columnwidth]{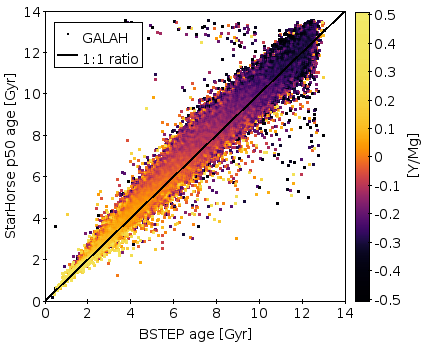}
    \caption{StarHorse vs astroNN ages in the left panel and Starhorse vs BSTEP ages colour coded by [Y/Mg] in the right panel.}
    \label{starhorse_apo_gal}
\end{figure}

We have compared our GALAH and APOGEE samples (without RGB and MS cuts) with StarHorse ages, which uses a isochrone-based bayesian estimation similar to BSTEP (Figure \ref{starhorse_apo_gal}), hence there is a much higher correlation between StarHorse x BSTEP ($r_{Pearson}=0.99$) ages than StarHorse x astroNN ages ($r_{Pearson}=0.67$).
In both APOGEE and GALAH, the StarHorse stars with available ages are concentrated in the turnoff region of the HR diagram, as shown in Figure \ref{HR_starhorse_k2}.

\begin{figure}
	\includegraphics[width=0.49\columnwidth]{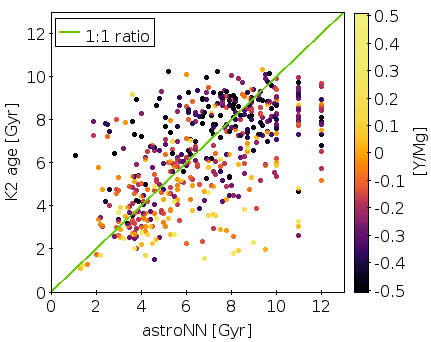}
	\includegraphics[width=0.49\columnwidth]{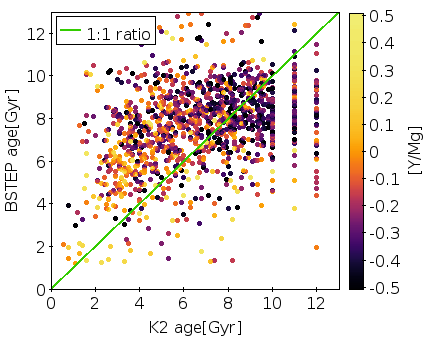}
    \caption{K2 vs astroNN ages for the common stars in the left panel and BSTEP vs K2 ages in the right panel, both colour-coded by [Y/Mg].}
    \label{k2_apo_gal}
\end{figure}

\begin{figure}
	\includegraphics[width=0.49\columnwidth]{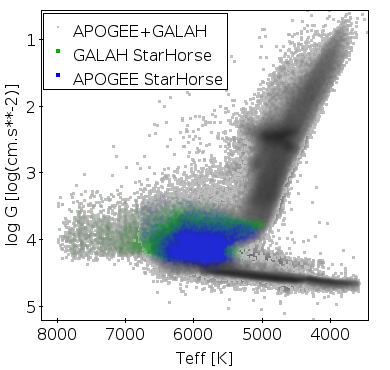}
	\includegraphics[width=0.49\columnwidth]{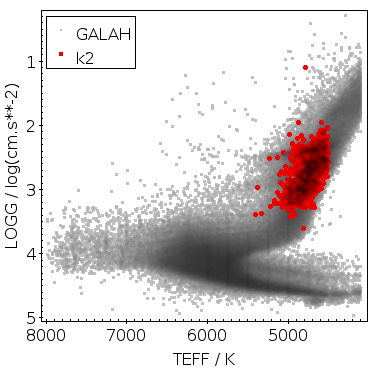}
    \caption{Sub-samples of stars with external age estimations and their locuses in the HR diagram. StarHorse stars with available ages are shown in the left panel and K2 stars are shown in the right panel.}
    \label{HR_starhorse_k2}
\end{figure}

In the same line, we also show a comparison of the ages obtained from asteroseismology in the K2 survey for a subsample of GALAH \citep{Zinn_2022} with our common sample (Figure \ref{k2_apo_gal}), where it shows a higher correlation with astroNN ($r_{Pearson}=0.57$) than with BSTEP ($r_{Pearson}=0.35$).
The K2 sample is concentrated in the red clump region of the HR diagram, also on Figure \ref{HR_starhorse_k2}.
The knee pattern seems to arise from all the asteroseismologic age to isochrone-based age relations present in this analysis.

It is clear that the different age estimation methods can lead to large biases and deviations if directly compared, and this is one of the main reasons that the age was not an input parameter for the t-SNE convergence.
Therefore, we only recommend comparisons between ages from the same estimation methods.

\begin{figure}
\centering
  \includegraphics[width=0.49\columnwidth]{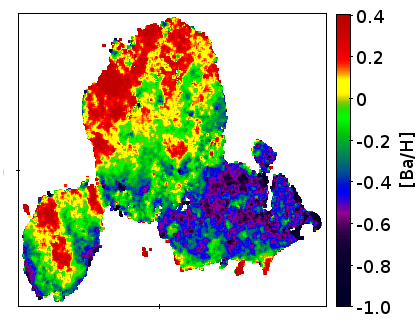}
  \includegraphics[width=0.49\columnwidth]{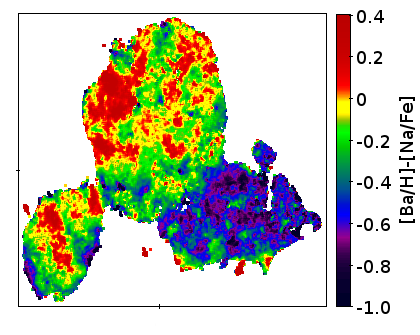} \hfill
  \includegraphics[width=0.49\columnwidth]{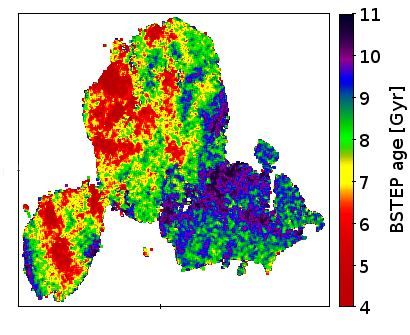}
  \includegraphics[width=0.49\columnwidth]{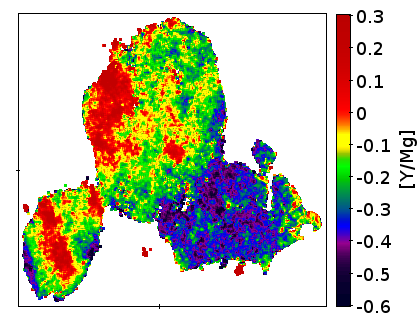} \hfill
  \includegraphics[width=0.49\columnwidth]{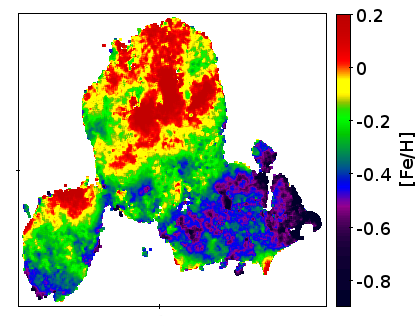}
  \includegraphics[width=0.49\columnwidth]{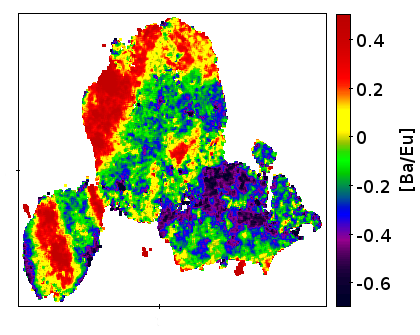} \hfill
  \caption{Preliminary t-SNE manifolds for the GALAH sample colour-coded by several abundance ratios and BSTEP ages. In these visualisations, the \textit{outer} disk (mostly R > 8 kpc) appears as the bottom-left blob, the \textit{inner} metal-rich disk is the large central concentration, and the dynamically hot alpha-enhanced population appears as the bottom-right concentration.}
\label{age_manifolds}
\end{figure}

During the convergence of t-SNE manifolds, we used an intermediate GALAH convergence to inspect several elemental ratios to compare their patterns with the BSTEP age estimation, as shown in Figure \ref{age_manifolds}.
The colours along different ratios were matched using TOPCAT's histogram-scaling to enable a fair and comprehensible pattern comparison and how the ratios behaved along the manifold.

In these manifolds, by qualitative comparison, we can see that the [Y/Mg] ratio can closely reproduce most of the features of the metal-rich moderate and younger age (< 8 Gyr) stars, but fails for the older metal-poor alpha-enhanced populations, underestimating the ages.
The [Fe/H] fails to reproduce the age relation in contrast with [Ba/H], which shows a very resemblant pattern to the age in the manifold.
We also found an improvement in the agreement for metal-rich stars by subtracting [Na/Fe] from [Ba/H].
The [Ba/Eu] relation showed the closest representation of the age pattern for the alpha-enhanced population, but not so well for the metal-rich portion. 
Also, APOGEE lacks a r-process element to normalise Ce, precluding a direct comparison with GALAH.

As cited by \cite{Kobayashi_2020}, light s-process elements are also produced by intermediate-mass asymptotic giant branch (AGB) stars, while heavy s-process elements are mainly from low-mass AGB stars, so a possible explanation for this direct [Ba/H]-age correlation is a continuous enrichment of Ba from low-mass stars along the evolution of the Galaxy, as the effects of type-Ia (main responsible for the Fe-peak enrichment, needing binary interacting systems) and type-II supernovae are not the main players in the Ba enrichment, with exception of the r-process contribution coming from massive stars, where the [Eu/Fe] can be used to account for this effect and filter the sample. 
Alternatively, [Na/Fe] can also be useful as a proxy to reduce this effect since Na can be produced either by hot-bottom-burning and carbon burning, both in massive stars \citep{Boeltzig_2016}.
Unfortunately, without any r-process elements in the APOGEE sample and the [Na/Fe] not properly measured for the metal-poor end (Figure \ref{Na_Al_comparison}), it leaves us with only Ce as an s-process element.
Since both Ba and Ce are heavy s-process elements, this enables the use of [Ce/H] as a possible auxiliary age-tracer in the APOGEE sample by comparing its behaviour with GALAH's [Ba/H] and age in the resulting manifolds presented.

It is good to remind that even though the chemical clocks may, in general, be good age estimators in the absence of more direct measurements such as asteroseismology, populations with anomalous abundances of heavy elements can lead to large mistakes when estimating their ages chemically.


\bsp	
\label{lastpage}
\end{document}